\documentclass[pdflatex,sn-mathphys-num]{sn-jnl}

\usepackage{graphicx}%
\usepackage{multirow}%
\usepackage{amsmath,amssymb,amsfonts}%
\usepackage{amsthm}%
\usepackage[title]{appendix}%
\usepackage{xcolor}%
\usepackage{textcomp}%
\usepackage{manyfoot}%
\usepackage{booktabs}%
\usepackage{algorithm}%
\usepackage{algorithmicx}%
\usepackage{algpseudocode}%
\usepackage{listings}%

\usepackage[T1]{fontenc}
\usepackage{lmodern}
\usepackage[mathscr]{eucal}
\hypersetup{bookmarksdepth=subsubsection}

	\definecolor{aaltoBlack}{RGB}{0,0,0}%
	\definecolor{aaltoGray}{RGB}{146,139,129}%
	\definecolor{aaltoRed}{RGB}{237,41,57}%
	\definecolor{aaltoBlue}{RGB}{0,101,189}%
	\definecolor{aaltoYellow}{RGB}{254,203,0}%
	\definecolor{aaltoPurple}{RGB}{102,57,183}%
	\definecolor{aaltoTurquoise}{RGB}{0,168,180}%
	\definecolor{aaltoGreen}{RGB}{0,155,58}%
	\definecolor{aaltoLightGreen}{RGB}{105,190,40}%
	\definecolor{aaltoOrange}{RGB}{255,121,0}%
	\definecolor{aaltoFuchsia}{RGB}{177,5,157}%

\begin{document}

\title[Article Title]{Leveraging active learning-enhanced machine-learned interatomic potential for efficient infrared spectra prediction}


\author[1,2]{\fnm{Nitik} \sur{Bhatia}}\email{nitik.bhatia@tum.de}

\author*[1,2,3,4]{\fnm{Patrick} \sur{Rinke}}\email{patrick.rinke@tum.de}

\author[2,5]{\fnm{Ond\v{r}ej} \sur{Krej\v{c}\'{i}}}\email{ondrej.krejci@utu.fi}

\affil*[1]{\orgdiv{Department of Physics}, \orgname{Technical University of Munich}, \orgaddress{\street{James-Franck-Strasse 1}, \city{Garching}, \postcode{85748}, \country{Germany}}}

\affil[2]{\orgdiv{Department of Applied Physics}, \orgname{Aalto University}, \orgaddress{\street{P.O. Box 11000}, \city{AALTO}, \postcode{FI-00076}, \country{Finland}}}

\affil[3]{\orgdiv{Atomistic Modelling Center}, \orgname{Munich Data Science Institute, Technical University of Munich}, \orgaddress{\street{Walther-Von-Dyck Str. 10}, \city{Garching}, \postcode{85748}, \country{Germany}}}

\affil[4]{\orgname{Munich Center for Machine Learning (MCML)}, \orgaddress{\city{Munich}, \country{Germany}}}

\affil[5]{\orgname{Department of Mechanical and Materials Engineering}, \orgaddress{\street{Vesilinnantie 5}, \city{Turku}, \country{Finland}}}



\abstract{
Infrared (IR) spectroscopy is a pivotal analytical tool as it provides real-time molecular insight into material structures and enables the observation of reaction intermediates \textit{in situ}. 
However, interpreting IR spectra often requires high-fidelity simulations, such as density functional theory based \textit{ab-initio} molecular dynamics, which are computationally expensive and therefore limited in the tractable system size and complexity. 
In this work, we present a novel active learning-based framework, implemented in the open-source software package PALIRS, for efficiently predicting the IR spectra of small catalytically relevant organic molecules.
PALIRS leverages active learning to train a machine-learned interatomic potential, which is then used for machine learning-assisted molecular dynamics simulations to calculate IR spectra.
PALIRS reproduces IR spectra computed with \textit{ab-initio} molecular dynamics accurately at a fraction of the computational cost. PALIRS further agrees well with available experimental data not only for IR peak positions but also for their amplitudes. 
This advancement with PALIRS enables high-throughput prediction of IR spectra, facilitating the exploration of larger and more intricate catalytic systems and aiding the identification of novel reaction pathways.
}




\maketitle

\section{Introduction}\label{intro}

Infrared (IR) spectroscopy has become an important analytical technique for the identification and characterization of chemical substances \cite{C3CS60374A, Khan2018}. Its applications span a multitude of disciplines, including chemistry, physics, biology, astrochemistry, astrophysics, and material sciences, where precise structural characterization is essential.

IR spectroscopy has been widely utilized to investigate gas-phase molecules, liquids, crystals, semicrystalline materials, amorphous solids, and interfaces such as solid/liquid, solid/solid, and solid/gas \cite{Rijs2015, COSTA2015449, Hamamoto2015, Tang2016, Griffith}. In catalysis, IR spectroscopy is particularly advantageous for probing reactions \textit{in situ}, enabling the identification of reaction intermediates and active sites, thus providing insight that facilitate the development of more efficient catalysts \cite{su_situ_2021, Cat_IR_2022}. Through these diverse applications, IR spectroscopy not only characterizes material structures but also elucidates the relationships between microscopic structures and macroscopic properties, thereby advancing both fundamental and applied research \cite{Peter2015, lansford_infrared_2020}.
 
However, the interpretation of experimental IR spectra and the accurate identification of molecular species remain challenging due to peak shifts and intensity variations induced by interactions with neighboring species, as well as \textit{spectral congestion} (also known as \textit{interference}) arising from overlapping vibrational signals \cite{Interpret_exp_IR, Intepret_exp_IR_2}. Consequently, theoretical calculations using quantum mechanical methods are often necessary to gain deeper insight and to ensure reliable interpretations \cite{Gaigeot2003, RODRIGUEZBETANCOURTT20111, Weymuth2012MOVIPACVS, AIMD_paper, Gaigeot2015, miotto_fast_2024, bastonero_automated_2024}.

The \emph{harmonic approximation} is the simplest way to compute IR spectra. In the harmonic approximation, vibrational frequencies are given by the second derivative of the potential energy, e.g., computed by DFT. Although effective, the harmonic approximation omits anharmonic effects that are important for accurate peak positions and the spectral shape \cite{AIMD_paper}.  \textit{Ab-initio} molecular dynamics (AIMD), in which changes of the molecular dipole moment are recorded over time, overcomes the limitations of the harmonic approximation \cite{AIMD_paper, Gaigeot2015}.  However, AIMD is computationally demanding, because long molecular dynamics (MD) trajectories are needed for accurate predictions of the peak positions and their amplitudes. Methods that can predict IR spectra accurately and efficiently are therefore urgently required.
 
Recent advancements in machine-learned interatomic potentials (MLIPs) make MLIPs a strong contender for accelerating or entirely replacing AIMD IR calculations. MLIPs learn the potential energy and interatomic forces from quantum mechanical calculations, typically DFT \cite{Behler_2007, bartok_gaussian_2010, smith_ani-1_2017, gastegger_machine_2017, SchNet, GM_2020, GM_2021, GPR, NequiP, Allegro}. If sufficiently trained, MLIPs provide accurate energies and forces, and thus MD trajectories, and combined with recent additions of dipole moment predictions \cite{grisafi_symmetry-adapted_2018, PhysNet, gastegger_machine_2021, PaiNN, beckmann_infrared_2022, Batatia2022Design, MACE} unlock IR spectra calculations three orders of magnitude faster than AIMD simulations~\cite{gastegger_machine_2021, tang_machine_2023, zou_deep_2023,stienstra_graphormer-ir_2024, krzyzanowski_machine_2024, yuan2025qme14scomprehensiveefficientspectral}.

However, the development of accurate MLIP-based frameworks for IR spectra predictions requires high-quality training datasets. Constructing such datasets is one of the most time- and cost-intensive aspects of MLIP development, as the data must capture the relevant interatomic interactions while its generation must not exceed computational budgets \cite{General_MLIP_1, bartok2018machine, deringer_general-purpose_2020}. Conventional data generation methods often involve exhaustive sampling, leading to large datasets with redundant information, increasing computational costs without necessarily improving model performance.
One promising solution to this challenge is active learning, a method that enables the systematic selection of the most informative data points, thus reducing the computational burden associated with both data generation and training \cite{AL_1, AL_2, Flare_1, van_der_oord_hyperactive_2023, AL_uncertainty-driven_2023, tang_machine_2023, tan_single-model_2023, zaverkin_uncertainty-biased_2024, Ghosh/etal:2025, Homm/Laakso/Rinke:2025}. By strategically focusing on regions of chemical space where the model's uncertainty is highest, active learning ensures that the data collected enhances the MLIP's accuracy and reliability while minimizing redundancy and inefficiencies.

Building on these advancements, we introduce PALIRS (A Python-based Active Learning Code for Infrared Spectroscopy), an active learning framework designed to efficiently construct training datasets for MLIP-based IR spectrum prediction. 
In this study, we first assess the effectiveness of PALIRS in training MLIPs-based on neural networks (NNs), and compare its capability to explore the configurational space with that of AIMD. Subsequently, we seek to evaluate PALIRS's ability to generate accurate IR spectra for small organic molecules relevant to catalysis. We assess how well ML-generated spectra compare to AIMD and experimental references, and determine the simulation time required for molecular dynamics-derived IR spectra to reach convergence. Additionally, we address the possibility to predict the temperature dependence of the predicted spectra, analyzing how spectral features evolve with temperature. Finally, we target the extrapolation limits of the full workflow by assessing its performance on molecules both similar and dissimilar to the training set. In doing so, we aim to quantify prediction errors in relation to molecular features and establish the boundaries of our ML model's generalizability.

\begin{figure}[t!]
	\centering 
    \includegraphics[width=1.0\textwidth]{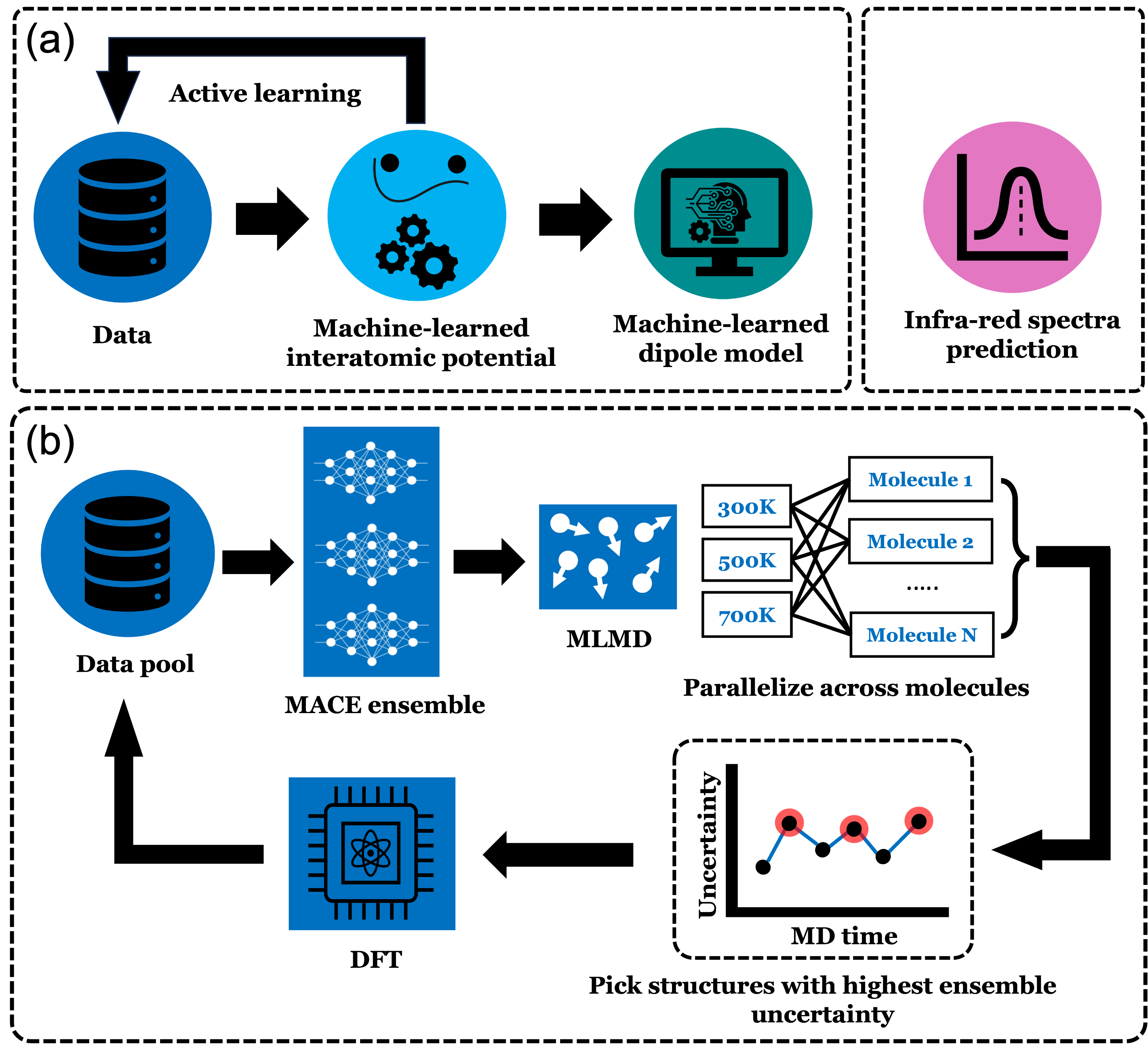}	
	\caption{
    \textbf{Workflow schematics for machine-learned infrared spectra prediction, as implemented in PALIRS:}
    (a) An overview of the machine learning methodology implemented for predicting infrared spectra from molecular structures. (b) The active learning framework designed to generate high-quality data for training machine-learned interatomic potential (MLIP) models, focusing on the construction of both potential energy surfaces and dipole moment surfaces. }
	\label{Combined_workflow}%
\end{figure}

\section{Results}\label{results}
\textbf{Computational workflow:}

To perform MD-based IR spectra predictions with MLIPs, we need two main ingredients: 1) MLIPs for accurate machine-learning-assisted molecular dynamics (MLMD) simulations and 2) accurate dipole moment predictions for computing the autocorrelation function and thus the IR spectrum. Therefore, an accurate description of energies, forces, and dipole moments is critical.

In this study, we introduce a four-step approach to predict the IR spectra of 24 small organic molecules (see Figure S1 in the Supplementary Information). This workflow is shown in Figure \ref{Combined_workflow}(a) and consists of the following steps:
(1) With the chosen molecules, we prepare an initial dataset of geometries to train the first version of the MLIP, which is then gradually improved through our active learning scheme.  After obtaining the final MLIP and dataset, (2) we train an additional ML model specifically to predict the dipole moments for each structure. (3) Using the MLIP for energies and forces, we proceed with MLMD production runs, and calculate dipole moments for all structures along the trajectory with the ML model. Finally, (4) the IR spectra are derived by computing the autocorrelation function of the dipole moments. These key steps are described in detail below. Further information on the implementation in the PALIRS package~\cite{PALIRS_repo} and specific settings are provided in the Methods section \ref{method}.

Our four-step approach (Figure \ref{Combined_workflow}(a)) works in principle with a single MLIP that provides an intrinsic estimate of uncertainty, such as GAP~\cite{bartok_gaussian_2010}.
The uncertainty estimation is a key feature of PALIRS'  active learning strategy.
Since NN-based MLIPs like MACE~\cite{Batatia2022Design, MACE} do not come equipped with an intrinsic uncertainty quantification, we employ an ensemble of three MACE models to approximate the uncertainty~\cite{schran_committee_2020, QBC_2}. 

\begin{figure}[t!]
	\centering 
    \includegraphics[width=0.98\textwidth]{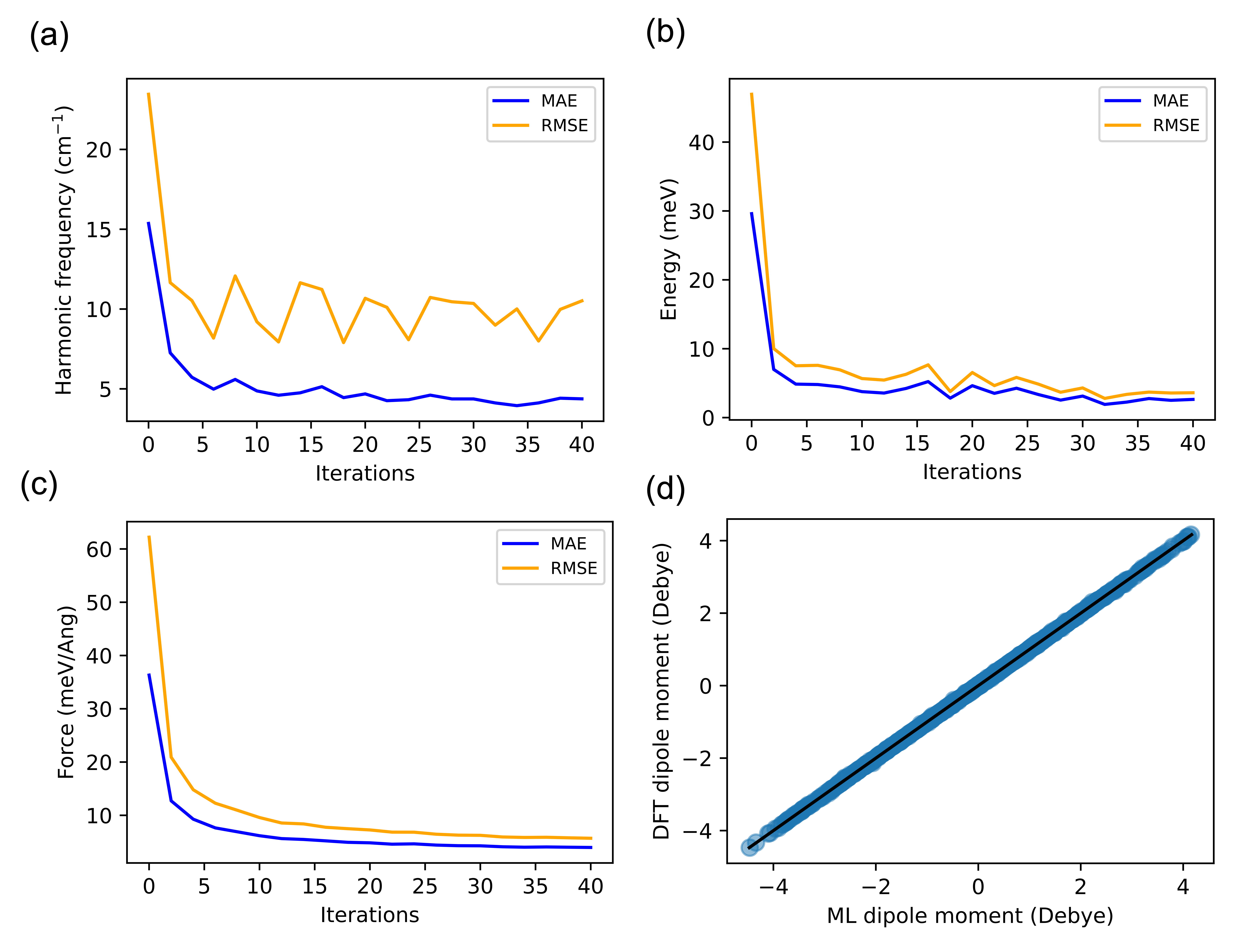}	
    \caption{\textbf{Assessment of MLIP and dipole ML model training and accuracy:} (a-c) Performance of the trained MLIP model, during the active learning procedure: (a) Harmonic frequency, (b) Total energy, (c) Force, evaluated using mean absolute errors (MAEs) and root mean squared errors (RMSEs). The harmonic frequency analysis is based on comparison with all 24 organic molecules, while energy and force evaluations are performed on an independent test set not used in training.
    (d) Accuracy of the dipole ML model the same test data as in (b) and (c), ensuring consistent performance evaluation across multiple properties.
    }
	\label{learning_curves}%
\end{figure}

The initial MACE MLIPs are trained on molecular geometries sampled along the normal vibrational modes \cite{AIMD_paper, smith_ani-1_2017, tang_machine_2023, qu_permutationally_2018} of each molecule. These geometries are obtained from DFT calculations performed with the FHI-aims code \cite{blum2009ab, havu2009efficient, levchenko2015hybrid,Xinguo/implem_full_author_list}. While these preliminary MLIPs provide a foundational representation of energies, and forces, their accuracy remains limited due to the relatively small training set, initially consisting of only 2085 structures. The modest performance of these initial models is evident in the learning curve shown in Figure~\ref{learning_curves} and will be further discussed in the next section.

To systematically refine the MLIP, we employ an active learning strategy (Figure \ref{Combined_workflow}(b)) that iteratively expands the training set through MLMD simulations. The acquisition strategy selects molecular configurations with the highest uncertainty in force predictions from each MLMD run, ensuring that the dataset is enriched with the most informative structures while minimizing redundancy. To balance exploration and exploitation during acquisition, the MLMD simulations are performed at three different temperatures: 300 K (low), 500 K (medium), and 700 K (high).
Further details on the active learning procedure are provided in the Methods section \ref{method}. The final dataset, after 40 active learning iterations, consists of 16,067 structures, with approximately 600–800 structures per molecule.

It is important to note that the active learning scheme focuses on optimizing energy and force predictions in the MLIPs. A separate ML model, also based on the MACE framework, is specifically trained to predict dipole moments for IR spectra calculations. We refer to this model as the dipole moment model in the following.

\textbf{Assessment of active learning performance:}
To assess MLIP improvement during active learning, we compare its predictions against a predefined test set of harmonic frequencies. These frequencies were obtained as a by-product of the normal mode sampling and include all 24 organic molecules of our study.
Harmonic frequencies serve as efficient validation, as they can be rapidly computed with MLIPs and directly compared to DFT reference values. Quantitative metrics such as the mean absolute error between MLIP- and DFT-computed harmonic frequencies provide a reliable measure of the model's accuracy and its progress in describing the studied molecules~\cite{tang_machine_2023}.

At each iteration, we used the first MLIP in the ensemble to evaluate the harmonic frequencies and quantified its accuracy using the mean absolute error (MAE) and root mean squared error (RMSE), as shown in Figure~\ref{learning_curves}(a).
The initial model, trained solely on molecular configurations from normal mode sampling, starts with an MAE of 15.36 and an RMSE of 23.45 cm$^{-1}$. As active learning progresses, these errors decrease, demonstrating the improvement of the MLIP. After approximately 30 iterations, the MAE reaches a plateau, indicating that the model could no longer be improved by adding more data using the current sampling strategy.  We then stop the active learning cycle with a final MAE of 4.37 and RMSE of 10.51 cm$^{-1}$ for the harmonic frequencies. 

To evaluate the performance of the MLIP in predicting energies and forces, we constructed a new test dataset generated from a 100 ps MLMD run at 300 K, using the first MLIP model from the final (40\textsuperscript{th}) iteration. Further details are provided in the Methods section.
This approach enables a retrospective analysis of how the accuracy of the MLIP evolves throughout the active learning process.
The corresponding MAEs and RMSEs are presented in Figures~\ref{learning_curves}(b) and~\ref{learning_curves}(c), showing a steady decrease in error with each iteration. In the final iteration, the MLIP achieves a MAE of 2.64 meV and an RMSE of 3.61 meV for energy predictions, and a MAE of 3.96 meV/Å with an RMSE of 5.69 meV/Å for force predictions. These low errors highlight the significant improvement in model accuracy as additional data is incorporated during training.

Subsequently, the dipole ML model is trained on the final dataset obtained through the active learning workflow. The model's accuracy, validated on the same test dataset used for the energy and force evaluations, is shown in Figure~\ref{learning_curves}(d).
The ML model demonstrates strong predictive performance, achieving an MAE of 7.62 and an RMSE of 12.46 mDebye. These results, along with the corresponding errors for energy and force predictions, are summarized in Table~\ref{table1}.

\begin{table}[htbp]
\caption{Final ML model performance on test data. MAE and RMSE stands for mean absolute error and root mean squared error, respectively.}\label{table1}%
\begin{tabular*}{0.5\textwidth}{@{\extracolsep\fill}lcccc}
\toprule
\multirow{2}{*}{Property} & \multirow{2}{*}{Unit} & \multicolumn{2}{c}{Test data} \\
\cmidrule{3-4}
                          &       & MAE   & RMSE \\
\midrule
Energy                    & meV   & 2.64  & 3.61  \\
Force                     & meV/Ang & 3.96  & 5.69  \\
Dipole moment             & mDebye & 7.62  & 12.46 \\
\botrule
\end{tabular*}
\end{table}

\begin{figure}[t!]
	\centering 
    \includegraphics[width=0.5\textwidth]{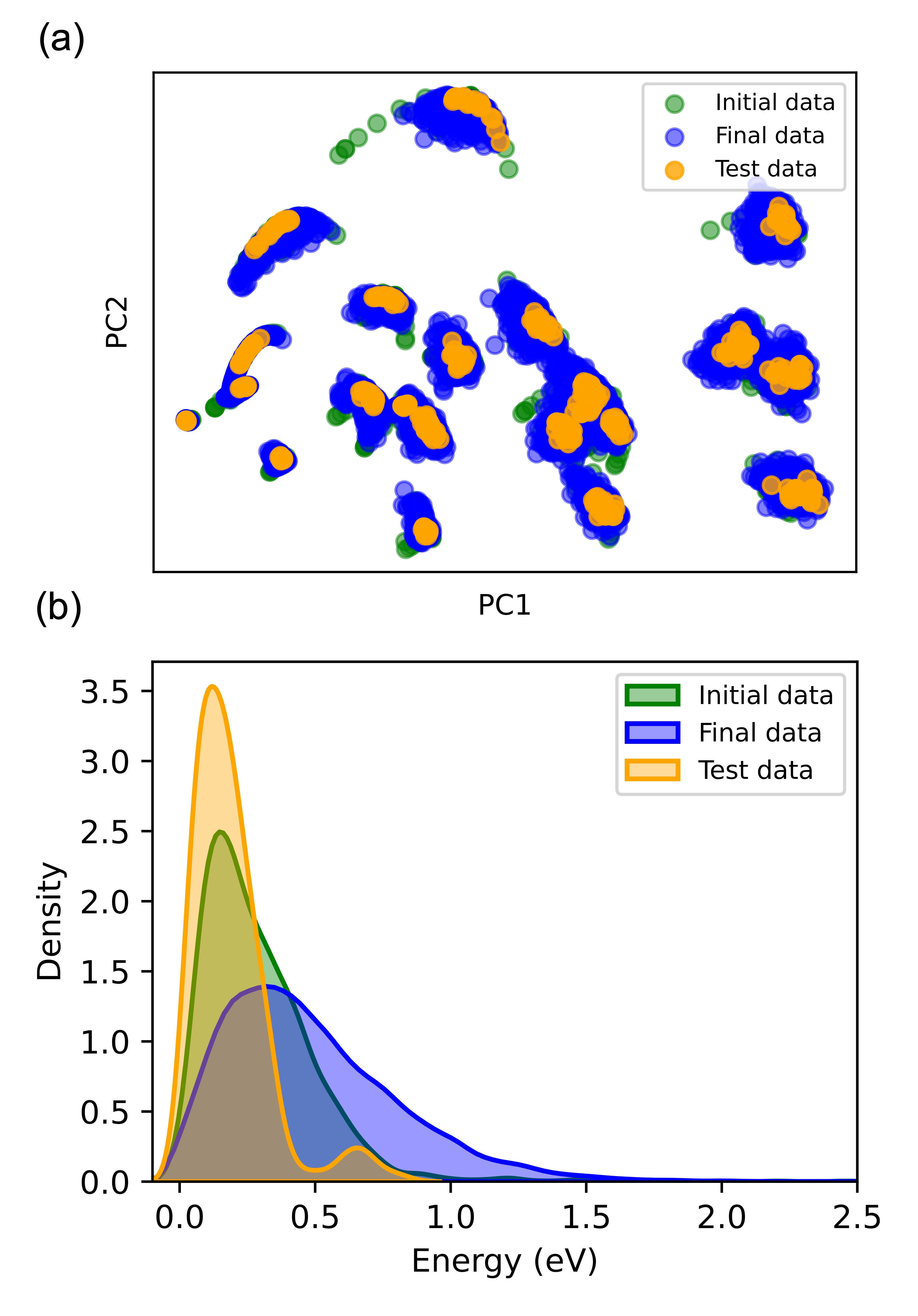}	
	\caption{\textbf{Data distribution analysis:} (a) Principal component analysis of the many-body tensor representation for all 24 organic molecules. (b) Energy distribution, with energies calculated by subtracting the optimized structure's energy. The three data categories are: Initial data (2085 samples, generated via normal mode sampling), final data (16,067 samples, obtained through active learning), and test data (480 samples, produced using the final MLIP).
    }
	\label{data_vis}%
\end{figure}

\textbf{Configurational and energy space exploration through active learning:}
PALIRS demonstrates high efficiency, characterized by exceptionally low error rates, even when trained on significantly fewer data points than conventional AIMD-based datasets. To investigate the underlying reasons for this performance, we visualized the configurational space using principal component analysis (PCA) applied to the many-body tensor representation (MBTR) \cite{Huo_2022, himanen2020dscribe} (Figure \ref{data_vis}(a)), and examined the corresponding energy distribution (Figure \ref{data_vis}(b)).

The PCA plot shows that our dataset of 24 small organic molecules forms several tightly grouped clusters. While some clusters are distinctly separated, many are closely spaced, merging into larger groupings. This indicates that the molecules share structural or chemical similarities, facilitating the MLIP’s ability to learn their properties. The plot also distinguishes initial, training, and test data using different colors, illustrating how the training data gradually expands to cover a broader region of chemical space compared to the sparser initial dataset. This progression, driven by active learning, helps sample underrepresented configurations. The test data is well-distributed across the space, enabling a robust evaluation of model performance.

The energy coverage shown in Figure \ref{data_vis}(b) is defined as the difference between each configuration’s total 
energy and that of the corresponding optimized molecule. Initially, the dataset spans a narrow energy window of approximately 1 eV. However, through active learning, the final dataset expands to cover a broader range exceeding 1.5 eV, despite the relatively small number of data points. This wider distribution -- still centered near zero -- ensures that both low- and high-energy configurations are well-represented, contributing to a more robust and generalizable MLIP.

The test data, derived from an extended MLMD trajectory at 300 K, spans an energy range of about 1 eV, making it well-suited for evaluating the model’s performance across the relevant energy landscape for IR calculations. The broader energy coverage of the final training set ensures that MLMD simulations for IR spectra remain within the domain of the MLIP and dipole models.

Additionally, Figure S2 (Supplementary Information) compares the configurational space explored by a DFT-based AIMD trajectory for methanol with that sampled via active learning. The results demonstrate that our active learning approach captures a wider energy range and a more diverse configurational space, highlighting its effectiveness in sampling complex molecular environments.

\textbf{Infrared spectra calculation and length of dynamical simulation:}

To compute the IR spectrum, a molecular dynamics (MD) trajectory is generated at a chosen temperature. The spectrum is then obtained by evaluating the autocorrelation function of the time derivative of the dipole moment along the trajectory~\cite{AIMD_paper}. The precise formulation and methodological details are provided in the Methods section~\ref{method}. However, the reliability of this approach is strongly influenced by the length of the MD simulation, as shorter trajectories can result in noisy or inconsistent spectral features~\cite{AIMD_paper}. To address this, we began our study with a systematic analysis aimed at identifying the minimum trajectory length required to achieve spectral convergence.

\begin{figure}[!t]
	\centering 
	\includegraphics[width=0.5\textwidth]{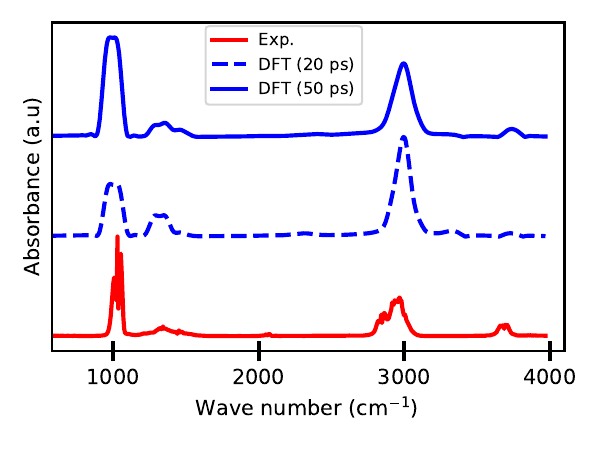}
    \caption{
    \textbf{The effect of MD run length on spectra prediction:}
    Comparison of gas-phase methanol spectra obtained by DFT-based AIMD simulations of 20 ps and 50 ps with the experimental spectrum (Exp.) obtained from the NIST database. All calculations and experiments were proceed at 300~K.
    }
    \label{AIMD_20_50}
\end{figure}

Figure~\ref{AIMD_20_50} presents the IR spectra of gas-phase methanol obtained from DFT-based AIMD simulations with two different trajectory lengths: 20 ps and 50 ps. The spectral peak positions converge by 20 ps, aligning with previous findings~\cite{AIMD_paper,esch_quantitative_2021}, suggesting that the key spectral features are already well captured at this timescale. However, the relative peak intensities at 20 ps exhibit an inverse trend compared to experimental spectra, and were also found to vary depending on the initial molecular geometry and velocity distribution. In contrast, the 50 ps simulation yields a spectrum that more closely matches the experimental NIST data~\cite{Wallace2024}. Notably, the experimental peak just above 1000 cm$^{-1}$ is significantly more intense than the one below 3000 cm$^{-1}$—a trend accurately reproduced in the 50 ps simulation. Based on these observations, we adopted 50 ps MD trajectories for all subsequent IR spectra calculations, including both DFT-based AIMD and ML-based predictions using MLMD simulations.

To quantify the similarity between the simulated IR spectra and the experimental data, we use Pearson's correlation coefficient (PCC) and the Wasserstein distance (WD), as these metrics have proven effective for assessing IR spectra similarity~\cite{esch_quantitative_2021}. PCC values range from -1 to 1 (with 1 indicating perfect similarity), while lower WD values indicate closer similarity, with 0 representing perfect alignment. A detailed description of these metrics is provided in the Methods section \ref{method}.

The PCC for the 20 ps simulation, shown in Figure~\ref{AIMD_20_50}, is 0.68, and increases to 0.73 for the 50 ps simulation. These values align well with previous DFT-based AIMD results~\cite{esch_quantitative_2021}. The improvement in PCC with longer simulations underscores the necessity of longer, more computationally demanding runs for accurate IR spectra predictions, reinforcing the efficiency of our approach implemented in PALIRS.

\begin{figure}[t!]
	\centering 
    \includegraphics[width=0.6\textwidth]{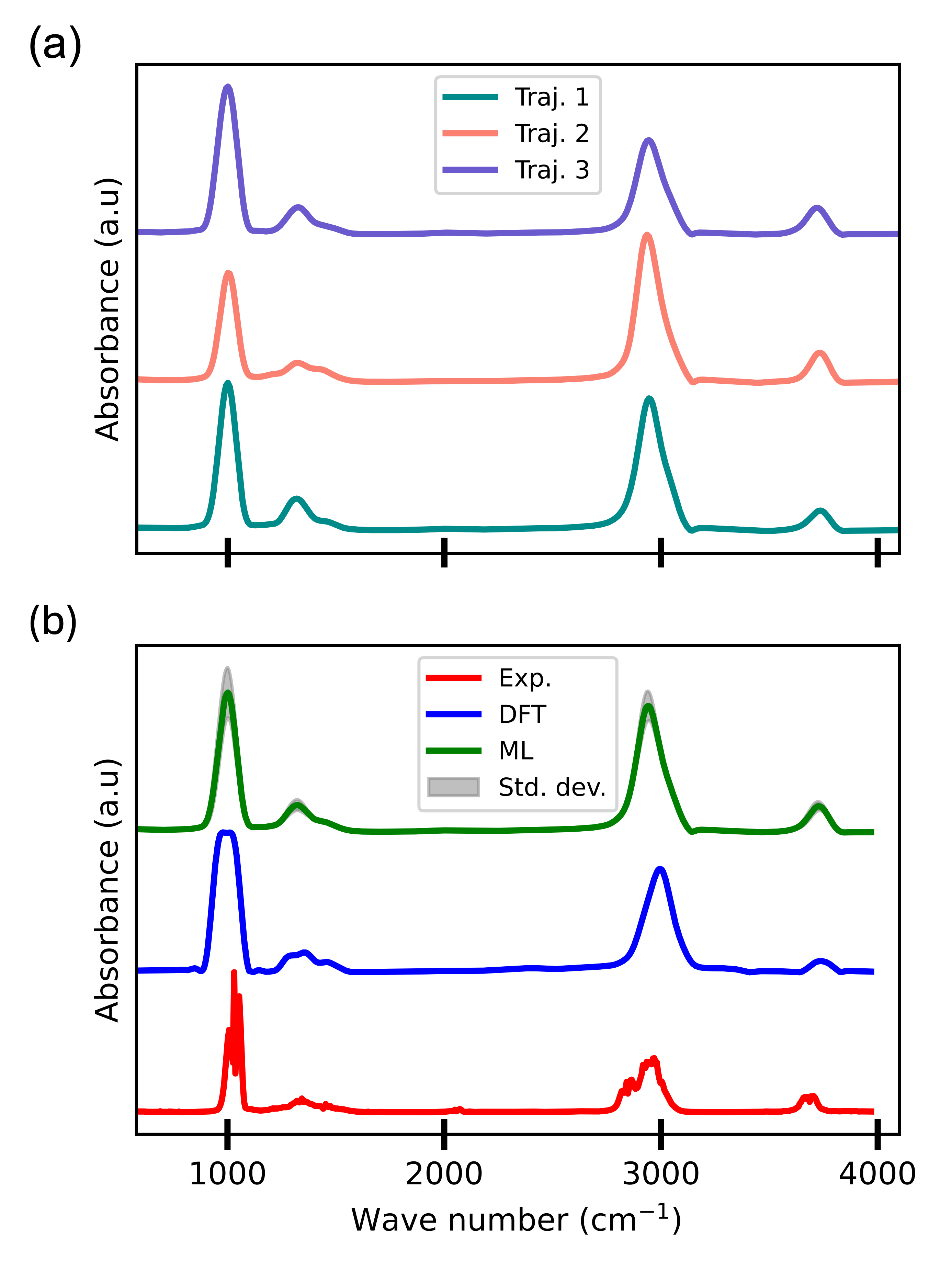}	
    \caption{\textbf{Infrared spectra of methanol in the gas phase at 300K:} (a) Spectra generated using three trajectories (Traj.) from the IR predicting ML models, and (b) a comparison between the experimental spectrum from the NIST database, DFT-based AIMD simulations, and the ML-predicted spectrum. averaged over three trajectories. The standard deviation (Std. dev.) of the ML spectrum is included to represent the uncertainty in the predictions.}
	\label{Methanol_spectra}%
\end{figure}

\textbf{Performance of PALIRS in predicting infrared spectra:}
With the simulation procedure established and the energy, force, and dipole moment accurately reproduced relative to DFT, we now proceed to the next critical step: evaluating the performance of the trained MLIP and dipole moment model in predicting IR spectra.
For the IR spectra prediction, we take advantage of the MLIP ensemble. Specifically, we conduct three separate MD simulations using the MLIP ensemble and the single dipole moment ML model (altogether referred to as ML models). 
For each MLMD trajectory, an IR spectrum is generated (Figure \ref{Methanol_spectra}(a)), and the final prediction is averaged across the three spectra. Additionally, the standard deviation between the runs is highlighted, providing an indication of the inherent uncertainty in the predictions.

Figure~\ref{Methanol_spectra}(b) presents a comprehensive comparison of the IR spectra for gas-phase methanol, including the ML-predicted spectrum, the experimental reference from the NIST database~\cite{Wallace2024}, and the DFT-based AIMD spectrum at 300 K. The MLMD and AIMD results show remarkable agreement, with both major and minor peaks aligning closely in terms of position and intensity. When compared to the experimental spectrum, the predicted intensities follow the same overall trend, with only minor shifts observed in peak positions.

\begin{table}[htbp]
\caption{Comparison of methanol spectra similarity across methods at 300K. PCC and WD are Pearson's correlation coefficient and Wasserstein distance, respectively.}\label{table_methanol_PCC_WD}%
\begin{tabular}{@{}lll@{}}
\toprule
Comparison & PCC  & WD \\
\midrule
Exp.-DFT    & 0.73 &	0.054    \\
DFT-ML    & 0.91 &	0.010    \\
Exp.-ML    & 0.80	& 0.057    \\
\botrule
\end{tabular}
\end{table}

The quantitative analysis in Table~\ref{table_methanol_PCC_WD} further confirms the minimal deviation between the DFT and ML results. Notably, it also indicates that the ML predictions align even more closely with the experimental data than the DFT-based AIMD results. This enhanced performance of the ML models can be attributed to the use of three independent MLMD simulations, which collectively sample a broader configurational space than a single DFT-based AIMD run. This more extensive sampling enables the ML-predicted IR spectrum to capture a wider range of molecular configurations, resulting in improved agreement with experimental observations. Overall, the ML models demonstrate a high degree of fidelity in reproducing spectral features, closely matching both experimental and DFT-based spectra.

The computed IR spectra for the remaining 23 organic molecules are available in the PALIRS repository~\cite{PALIRS_repo}, while the corresponding PCCs and WDs are provided in the Supplementary Information (Figure S3). A summary of the statistical analysis is presented in Table~\ref{avg_PCC_WD_table}.

Overall, the ML model exhibits excellent accuracy in predicting IR spectra, showing strong agreement with both DFT-based and experimental results. This underscores the efficiency of our approach in significantly reducing computational cost without compromising predictive performance. Leveraging our active learning strategy, fewer than 1,000 single-point DFT calculations per molecule were sufficient to generate accurate IR spectra—compared to the over 100,000 required for DFT-based AIMD.

For example, a 50 ps MLMD simulation of methanol produced results comparable to DFT-based AIMD in just 1 hour on an NVIDIA Volta V100 GPU, whereas the DFT-based simulation required 107 CPU hours per core on an Intel Xeon Gold 6230. This dramatic speedup not only benefits small molecules like methanol but also scales favorably for larger systems. 
While MLMD runtimes scale linearly with the number of atoms \(N\), i.e., \(\mathcal{O}(N)\), the computational cost of DFT-based AIMD increases approximately as \(\mathcal{O}(N^3)\). This stark contrast makes MLMD a highly scalable and efficient alternative for simulating larger systems.

\begin{table}[htbp]
\caption{Quantitative estimation of spectral similarity across all the 24 organic molecules. The experimental data (Exp.) were obtained from the NIST database. The line above a quantity indicates the mean, while $\delta$ represents the standard deviation.}\label{avg_PCC_WD_table}%
\begin{tabular}{@{}lllll@{}}
\toprule
Comparison & $\overline{\mathrm{PCC}}$ & $\delta$PCC & $\overline{\mathrm{WD}}$ & $\delta$WD\\
\midrule
Exp. - DFT    & 0.68   & 0.22   & 0.045   & 0.025  \\
DFT - ML    & 0.80   & 0.18   & 0.026   & 0.022  \\
Exp. - ML    & 0.81   & 0.15  & 0.029   & 0.013  \\
\botrule
\end{tabular}
\end{table}

\begin{figure}[t!]
	\centering 
    \includegraphics[width=0.6\textwidth]{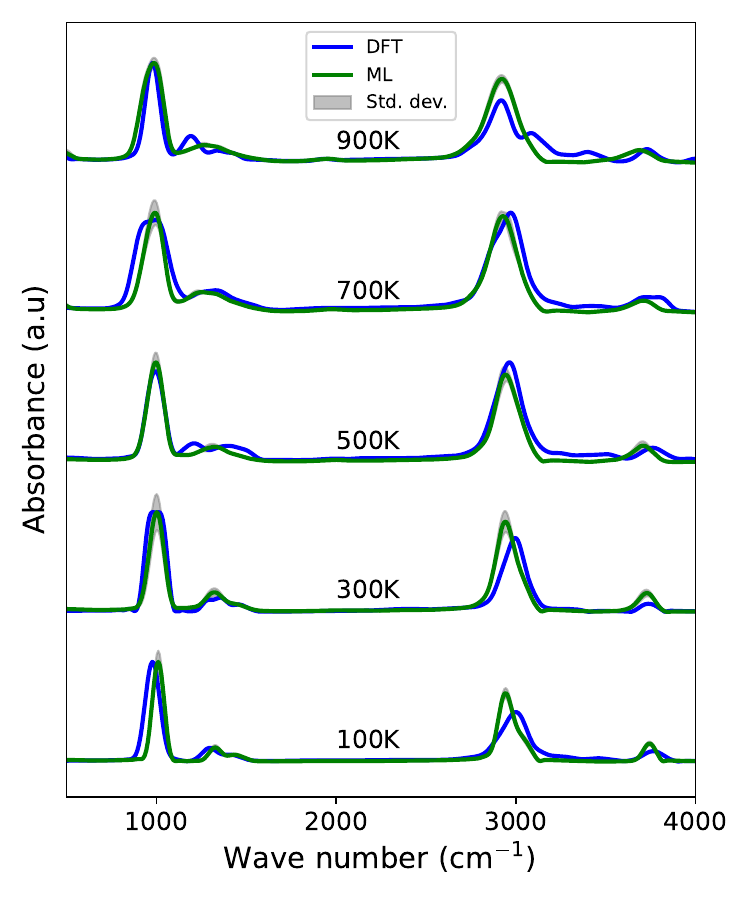}	
    \caption{\textbf{Assessment of the ML model temperature dependence:} Infrared spectra of methanol in the gas-phase at five different temperatures (100K, 300K, 500K, 700K and 900K) calculated with the ML models and DFT-based AIMD.}
	\label{Methanol_temp}%
\end{figure}

\textbf{ML model performance across temperature:}
To further test our IR prediction capabilities, we compare the DFT-based AIMD spectra of methanol with the ML-predicted spectra at five different temperatures in Figure \ref{Methanol_temp}.
At both low and high temperatures, the band positions are well predicted by the ML models. However, the intensity around 3000 cm$^{-1}$ is slightly overestimated compared to the DFT-based AIMD results. The average PCC and WD values for different temperatures are 0.93 and 0.003, respectively, with standard deviations of 0.046 and 0.001. For detailed values, please refer to the Supplementary Information (Figure S4).
Overall, the results clearly indicate that the ML models have effectively captured the temperature-dependent behavior of the spectra, demonstrating accurate predictions across a wide range of temperatures. An identical analysis has been conducted for ethanol in the gas phase, with the calculated IR spectra and corresponding PCC and WD values also available in the Supplementary Information (Figure S5), showing even better similarity between the ML and DFT-predicted spectra.

\textbf{Assessment of ML model transferability:}
To evaluate the ML models' performance beyond the organic species included in the training data, we selected a set of 8 molecules with increasing carbon counts and varying functional groups. Using the same procedure, we compare the average ML-predicted spectra with the experimental data from the NIST database. The full set of 8 molecules and their PCCs and WDs are provided in Figure~S6 in the Supplementary Information, offering a quantitative assessment of the ML models' transferability.

\begin{figure}[t!]
	\centering 
    \includegraphics[width=0.6\textwidth]{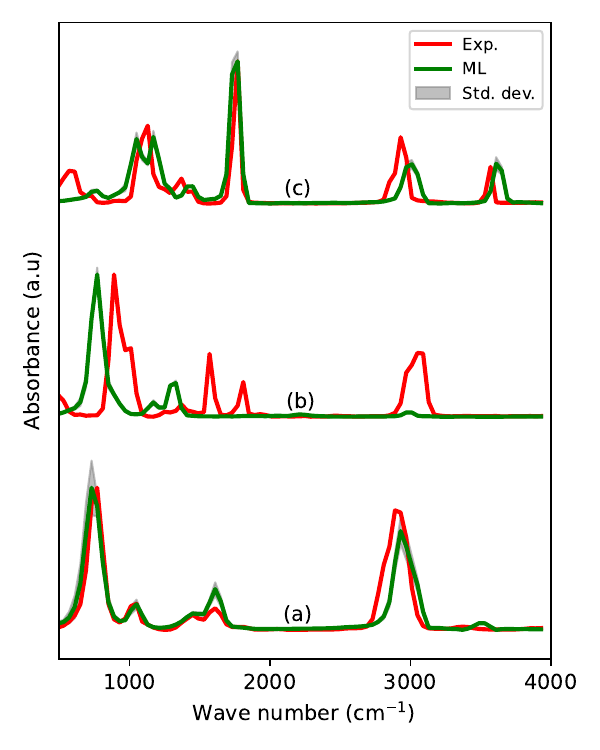}	
    \caption{\textbf{Transferability of ML models in the prediction of infrared spectra in gas-phase:} (a) Methyl amine, (b) 1,3-butadiene, (c) pentanoic acid.}
	\label{Transferable_spectra}%
\end{figure}

In Figure~\ref{Transferable_spectra}, we present three representative cases illustrating the ML models' best, worst, and intermediate performance. The best agreement is observed for methyl amine, where both peak positions and intensities closely match the experimental spectrum. In contrast, for 1,3-butadiene, significant deviations in both frequency and intensity patterns are evident. Pentanoic acid shows intermediate agreement, with reasonable alignment in peak positions and as well as in relative intensities.
These results highlight a clear dependence of predictive accuracy on the presence of similar chemical environments in the training data, as reflected in the quality of the predicted spectral features.

\section{Discussion}\label{sec3}
The MLIPs and the dipole moment ML model demonstrate strong potential for accelerating IR spectrum prediction of small molecules. 
Our results indicate that our active learning strategy reduces the amount of data needed per molecule by a factor of 10 compared to earlier ML-based approaches~\cite{PhysNet, gastegger_machine_2021, PaiNN, beckmann_infrared_2022}, and by a factor of 100 compared to conventional AIMD methods, while maintaining good agreement with AIMD reference spectra. Moreover, the training configurations selected by PALIRS effectively cover the configuration space (Figure \ref{data_vis}), underscoring the ability of active learning to focus on physically meaningful variations rather than redundant data points.

Building on these results, the strong qualitative agreement between PALIRS predictions and reference spectra suggests that ML-based models can implicitly capture anharmonic effects—phenomena that are typically difficult and computationally expensive to model using conventional ab initio methods. In contrast to explicit anharmonic approaches such as the vibrational self-consistent field (VSCF) method, its perturbative extensions, second-order vibrational perturbation theory (VPT2), or vibrational configuration interaction (VCI)—all of which scale poorly with system size~\cite{sagiv_anharmonic_2018}—PALIRS offers a highly efficient alternative.

By leveraging MLMD, PALIRS enables spectral convergence through longer trajectories at a fraction of the computational cost of AIMD. This efficiency makes it particularly well-suited for high-throughput evaluation of vibrational spectra, especially for catalytically relevant intermediates~\cite{PhysNet, gastegger_machine_2021, PaiNN, beckmann_infrared_2022, tang_machine_2023, zou_deep_2023, stienstra_graphormer-ir_2024, krzyzanowski_machine_2024, yuan2025qme14scomprehensiveefficientspectral}.

Given the temperature dependence of vibrational spectra, we assessed PALIRS at various temperatures and found it accurately captures temperature-induced shifts in peak positions and intensities. This highlights the ability of MLMD to capture temperature-dependent spectral changes without explicitly recalculating forces at each temperature using computationally expensive AIMD simulations. However, we also observe that temperature effects introduce subtle discrepancies, particularly in the broadening of certain peaks at higher temperatures. These differences suggest that while MLMD-based IR spectra effectively account for anharmonic effects, further improvements in training set diversity could enhance robustness at extreme conditions.

While PALIRS demonstrates strong performance for small organic molecules represented in the training set, its predictive accuracy diminishes for more complex or chemically distinct systems. For example, the poor performance observed for 1,3-butadiene can be attributed to the limited representation of C=C bonds in the training data—ethene being the only such example—restricting the model’s ability to generalize to similar bonding environments. In contrast, molecules like methyl amine and pentanoic acid, which contain nitrogen and oxygen functional groups that are well-represented in the training data, are predicted with reasonable accuracy. Enhancing the training set to incorporate a wider range of bonding motifs and chemical diversity would help improve generalization. In this context, transfer learning offers a promising pathway to adapt PALIRS for new molecular families.

The limitation in generalization is also reflected in the uncertainty of force predictions during MLMD simulations. For molecules within the training distribution, the maximum force uncertainty across the ensemble of MLIPs remains low, around $10^{-5}$ eV/Å. However, it increases to $10^{-3}$ eV/Å for methyl amine and reaches up to $10^{-1}$ eV/Å for 1,3-butadiene. Despite these elevated uncertainties, the predicted IR spectra remain consistent, indicating that variations in force predictions do not necessarily translate into spectral deviations. This is also evident from the standard deviations shown in Figure~\ref{Transferable_spectra}. Incorporating explicit uncertainty quantification into the spectral predictions could further strengthen the model’s robustness, particularly for out-of-distribution systems.

\section{Conclusion}\label{sec4}

We demonstrated the effectiveness of combining active learning with machine-learned interatomic potentials (MLIPs) and a dipole moment model for predicting IR spectra of small organic molecules. The PALIRS-trained models achieved accuracy comparable to or better than DFT-based AIMD, while requiring 100 times fewer DFT calculations. A 50 ps MD trajectory was found sufficient for spectral convergence, and the ML models accurately captured temperature-dependent features and generalized well to larger, structurally similar molecules. Given the relevance of many studied molecules to catalysis, this approach offers a scalable, cost-efficient alternative for high-throughput spectral analysis. Future work will expand the dataset via transfer learning to support inverse molecular design and deeper exploration of catalytic systems.

\section{Methods}\label{method}
\subsection{Machine learning models}

We trained the equivariant message-passing neural network, MACE \cite{Batatia2022Design, MACE} and used it to predict total energies and forces. We trained the MACE model with 128 channels, (L=1) equivariant messages, and a cutoff radius of 3.0 Å. The models were structured with two layers, each having a body order of 2, and were trained with MACE version 0.3.5. 

In addition, a MACE model was trained to predict dipole moments for infrared spectrum calculations. The model architecture mirrored that of the previously described MACE framework, with the key difference being that the readout function outputs a vector representing the dipole moment for each atom, rather than scalar atomic site energies.

All our MACE training and predictions were performed on NVIDIA Volta V100 GPUs.

\subsection{DFT computational details}
All DFT calculations were performed using the all-electron numeric-atom-centered orbital code FHI-aims \cite{blum2009ab, havu2009efficient, levchenko2015hybrid,Xinguo/implem_full_author_list}. The Perdew-Burke-Ernzerhof exchange-correlation functional (PBE) \cite{perdew_generalized_1997} was used for the calculations. Further computational settings included the standard FHI-aims tier-1 basis sets and "light" grid settings, the zeroth-order regular approximation to account for scalar relativistic effects \cite{lenthe1993relativistic}, and a Hirshfeld correction term for van der Waals interactions \cite{tkatchenko_accurate_2009}.
Structure optimization was carried out using the Broyden–Fletcher–Goldfarb–Shanno (BFGS) minimizer \cite{nocedal_numerical_2006}, with a convergence limit of 1 meV/Å for the maximum atomic force amplitude.
Geometry optimization and AIMD simulations were performed using the conventional FHI-aims software, while the normal mode sampling and active learning tasks were conducted using the Atomic Simulation Environment (ASE) \cite{ase-paper} and the FHI-aims calculator.

\subsection{Initial data generation and MLIP model}
Our dataset comprises 24 organic molecules containing hydrogen, carbon, nitrogen, and oxygen. Each molecule has up to two carbon atoms 
(Figure~S1). Their starting geometries were extracted from the QM9 dataset \cite{ramakrishnan_quantum_2014} and subsequently re-optimized at the DFT level. Starting from the optimized geometries, static calculations based on the harmonic oscillator approximation were performed, and the structures were sampled along the normal modes \cite{AIMD_paper, tang_machine_2023, smith_ani-1_2017, qu_permutationally_2018}. Each normal mode corresponds to a distinct way of displacing the structure from its local minimum energy configuration, with all atoms oscillating at the same frequency (refer to Figure~S7 in Supplementary Information)\cite{AIMD_paper, tang_machine_2023, smith_ani-1_2017}. We used the first 10 geometries from each of the normal modes sampled by ASE with its default settings~\cite{ase-paper}.

Single-point DFT calculations were performed on the sampled data to compute total energies, forces, and dipole moments. In the end, 2085 structures (50-200 structures per molecule) were collected and used to train the initial ensemble of MACE MLIP models to start the active learning workflow. The entire dataset was split into a training subset and a test subset with a ratio of 80:20 during initial and subsequent training. 
The MLIP ensemble was trained on the same dataset. 
The ensemble MLIPs differ only in the random \texttt{seed} parameter, which controls the initialization of model weights during MACE training.

\subsection{Active learning workflow}
The ensemble of MACE MLIPs enables the identification of structures with high uncertainties \cite{schran_committee_2020, QBC_2}, which is a key concept of our active learning workflow. The ensemble MLIPs are iteratively improved by augmenting the dataset with the 15 structures of each molecule with the highest uncertainty in each iteration. This selection process leverages three MLMD simulations conducted at 300, 500, and 700 K, extracting the five most uncertain structures from each temperature. The use of low, intermediate, and high temperatures strikes an optimal balance between exploring new regions of the chemical space and effectively exploiting known regions. A sanity check is implemented that monitors the forces of configurations during the MLMD simulations. For each configuration, we calculate a relative force error as the root of the variance of the predicted forces normalized by the force magnitudes, with a small regularization factor added to prevent division by zero. If this relative error exceeds a threshold of 0.5, the MD simulation is immediately terminated to avoid the generation and propagation of unreliable structures.
At each iteration of the active learning cycle, MLMD sampling is performed by continuing the trajectory from the previous iteration, using its final geometry and velocities as the initial conditions.

For the new structures selected by each iteration, we perform single point DFT calculations and add the results to our dataset. After splitting the updated dataset again, all three models are retrained from scratch with different random \texttt{seeds} to remove bias from previous training states, ensures diversity among the models in the committee, and to provide a more reliable estimate of the predictive uncertainty.
We repeat this cycle until the estimated error in harmonic frequencies either falls below the selected threshold (5 cm$^{-1}$) or reaches a maximum of 40 iterations.

To accelerate the sampling process on a high-performance computing cluster, the 24 molecules are divided into three batches, each containing 8 molecules. Within each batch, the molecules are sampled in parallel at different temperatures, optimizing the use of computational resources efficiently. Dividing into batches helps balance the workload and prevent bottlenecks in resource allocation, ensuring a more efficient and manageable distribution of computational tasks.

Finally, 600-800 instances of uncertain data are collected per molecule, and the final training data consisted of 16,067 structures. With the final dataset we train the  MACE dipole model. 

\subsection{PALIRS}
The PALIRS repository \cite{PALIRS_repo} provides all necessary scripts for initial data generation, setting up the active learning workflow, and the final trained models. It also includes detailed documentation and a concise tutorial to facilitate a seamless workflow setup. While PALIRS is configured for MACE MLIPs and FHI-aims, it can be readily adapted to other ASE-compatible MLIPs and DFT codes.

\subsection{Test set generation}
\begin{figure}[t!]
	\centering 
    \includegraphics[width=1.0\textwidth]{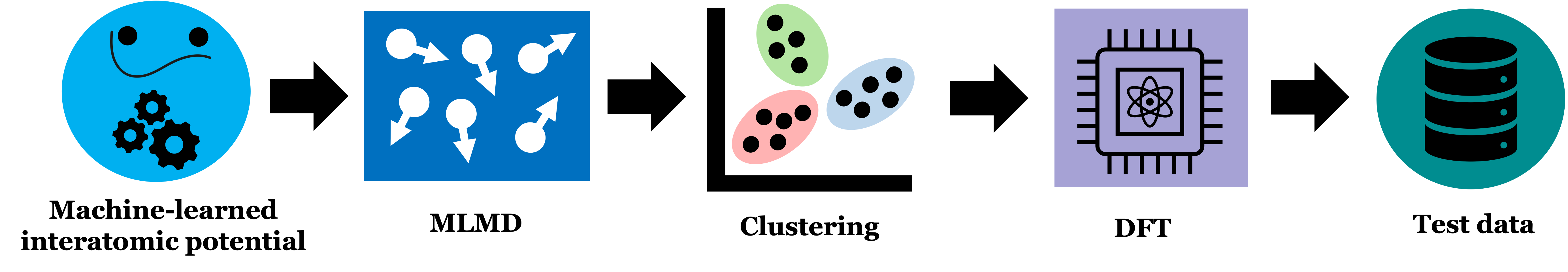}	
	\caption{\textbf{Test data:}
    A streamlined approach for constructing a test dataset, validating the ML models generated at each iteration of the active learning loop.
    }
	\label{Test_data}%
\end{figure}

To evaluate whether the final ensemble of MLIPs can be reliably applied to IR simulations, the first MLIP of the ensemble is used to create a test dataset, as illustrated in Figure~\ref{Test_data}.  
First, a 300 K, 100 ps MLMD simulation is performed for each molecule, from which 2000 atomic structures are uniformly sampled. These structures are then represented using the MBTR descriptor \cite{Huo_2022}, as implemented in the DScribe library \cite{himanen2020dscribe}. We use K-means clustering \cite{bennett2000constrained} in the MBTR feature space to select 5 clusters. 4 structures are randomly selected from each cluster, yielding a total of 20 structures per molecule.
This clustering-based selection effectively reduces bias and guarantees a broad coverage of the configurational space \cite{Homm/Laakso/Rinke:2025}. In total, the final test dataset consists of 480 structures, and DFT calculations are performed on the selected structures.

\subsection{MD simulations} 
DFT-based AIMD IR spectra calculations proceed in two stages: First, the system is equilibrated using a Berendsen thermostat with a relaxation time of 0.1 ps \cite{berendsen_molecular_1984}, at a specified temperature for 4 ps. Subsequently, a Nosé-Hoover thermostat, with an effective thermostat mass of 4000 cm$^{-1}$ \cite{nose_unified_1984, hoover_canonical_1985}, is employed for a 50 ps run. Only the Nosé-Hoover run is utilized for computing the IR spectra.

All MLMD runs are performed with a Langevin thermostat \cite{ceriotti_langevin_2009} with a friction coefficient of 0.01 as implemented in ASE 3.22.1 \cite{ase-paper}. 5 ps MLMD runs are used during active learning to identify highly uncertain structures. 
The ML-predicted IR spectra are obtained from three independent MLMD trajectories, each initiated with different velocity seeds and propagated using forces obtained by averaging predictions from a committee of models. First, 5 ps of each trajectory is used for thermalization, while the following 50 ps are used for the IR simulation.

For all MD simulations, a time step of 0.5 fs is used, and,  unless specified otherwise, the temperature is set to 300 K.

\subsection{Infrared spectra calculation}
The IR spectrum is calculated from an MD trajectory by computing the auto-correlation function of the time derivative of the dipole moment, $\dot{\mu}$, according to the following equation \cite{AIMD_paper}:

\begin{equation} I_{IR}(\omega) \propto \int_{-\infty }^{+\infty}\left\langle \dot{\mu}\left( \tau \right) \dot{\mu}\left( \tau + t \right) \right\rangle_{\tau}e^{-i\omega t}dt .\label{eq:IR-intensity} \end{equation}

\noindent
In this work, all auto-correlation functions are computed using the Wiener–Khinchin theorem \cite{wiener_generalized_1930}. A Hann window function \cite{blackman_measurement_1958} and zero-padding are applied to the auto-correlation functions before the Fourier transform to obtain high-quality IR spectra. A maximum correlation depth of 1000 fs is used. For IR spectra processing, we utilized the autocorrelation function from SchNetPack \cite{schutt_schnetpack_2019} and integrated it into our workflow.

\subsection{Spectra preprocessing and similarity measures}
To enable accurate comparison between experimental and theoretical IR spectra, it is crucial to correct for artifacts present in the data. A common issue in experimental spectra is baseline correction, which can obscure spectral features and hinder meaningful analysis. We address this by automating baseline correction using the method described in \cite{esch_quantitative_2021}, as illustrated in Figure~S8 of the Supplementary Information. Additionally, to reconcile differences in frequency sampling between experimental and theoretical spectra, we apply linear interpolation to map the theoretical data onto the experimental frequency grid.

For comparison between theories or between theory and experiment, we utilize two metrics to evaluate the degree of correlation between the spectra: Pearson's correlation coefficient (PCC) (Equation \ref{eq:PCC}) and the Wasserstein distance (WD) (Equation \ref{eq:WD}), both implemented in scikit-learn 1.5.2 \cite{scikit-learn}. The PCC is defined as:

\begin{equation}
\text{PCC} = \frac{\sum_{i=1}^{n} (x_i - \bar{x})(y_i - \bar{y})}{\sqrt{\sum_{i=1}^{n} (x_i - \bar{x})^2} \sqrt{\sum_{i=1}^{n} (y_i - \bar{y})^2}},\label{eq:PCC} \end{equation}
where $x_i$ and $y_i$ represent the intensities of the respective spectra, while $\bar{x}$ and $\bar{y}$ are their mean values \cite{esch_quantitative_2021, henschel_theoretical_2020, pracht_comprehensive_2020}.

The WD is defined in terms of an integral, rather than the discrete definition of PCC:

\begin{equation}
WD(\mu, \nu) = \inf_{\gamma \in \Gamma(\mu, \nu)} \int_{\mathbb{R} \times \mathbb{R}} ||x - y|| \ d\gamma(x,y),\label{eq:WD} \end{equation}
where $\mu$ and $\nu$ are the distributions corresponding to the two spectra being compared. $\Gamma(\mu, \nu)$ denotes the set of all joint distributions with marginals $\mu$ and $\nu$, $||x - y||$ is the distance between points sampled from the distributions, and $d\gamma(x,y)$ represents the infinitesimal mass transported from $x$ to $y$ \cite{esch_quantitative_2021, rubner_earth_2000}.

\backmatter

\bmhead{Supplementary information}
The Supplementary  information contains:~\\
Figure S1 -- A set of 24 representative small organic molecules used in this study.
Figure S2 -- Energy distribution of methanol, comparing DFT-based AIMD and AL data relative to the optimized structure.
Figure S3 -- Spectral similarity for 24 molecules, comparing Exp.-DFT, DFT-ML, and Exp.-ML using PCC and WD.
Figure S4 -- IR spectra of methanol at five temperatures, with DFT-ML similarity evaluated at each temperature using PCC and WD.
Figure S5 -- IR spectra of ethanol at five temperatures, with DFT-ML similarity evaluated at each temperature using PCC and WD.
Figure S6 -- Spectral similarity for eight organic molecules, comparing Exp.-ML spectra using PCC and WD.
Figure S7-- Visualization of normal modes of methanol and sampling of structures along them.
Figure S8 -- Baseline correction applied to the experimental spectrum, showing improved alignment with ML predictions.

\bmhead{Acknowledgments}
N.B. acknowledges the funding from Horizon Europe MSCA Doctoral network grant n.101073486, EUSpecLab, funded by the European Union.
O.K. and P.R. have received funding from the European Union – NextGenerationEU instrument and are funded by the Research Council of Finland (grant numbers 348179, 346377, and 364227). 
We acknowledge CSC, Finland for awarding access to the LUMI supercomputer, owned by the EuroHPC Joint Undertaking, hosted by CSC (Finland) and the LUMI consortium through CSC, Finland, extreme-scale project ALVS.
The authors also gratefully acknowledge the additional computational resources provided by CSC – IT Center for Science, Finland, and the Aalto Science-IT project.

\section*{Declarations}
The authors report no conflicts of interest. To facilitate open materials science \cite{Himanen/Geurts/Foster/Rinke:2019}, the reference DFT-based AIMD dataset is accessible at 10.5281/zenodo.14657903, while the data curated through the active learning scheme is available at 10.5281/zenodo.14699673. Additionaly, the spectral data published in this study can be found in~\cite{PALIRS_repo}.

N.B. created the workflow and proceeded with the calculation.
O.K. and P.R. supervised the work.
All authors contributed to the manuscript.

\bibliography{sn-bibliography}


\begin{thebibliography}{86}
\ifx \bisbn   \undefined \def \bisbn  #1{ISBN #1}\fi
\ifx \binits  \undefined \def \binits#1{#1}\fi
\ifx \bauthor  \undefined \def \bauthor#1{#1}\fi
\ifx \batitle  \undefined \def \batitle#1{#1}\fi
\ifx \bjtitle  \undefined \def \bjtitle#1{#1}\fi
\ifx \bvolume  \undefined \def \bvolume#1{\textbf{#1}}\fi
\ifx \byear  \undefined \def \byear#1{#1}\fi
\ifx \bissue  \undefined \def \bissue#1{#1}\fi
\ifx \bfpage  \undefined \def \bfpage#1{#1}\fi
\ifx \blpage  \undefined \def \blpage #1{#1}\fi
\ifx \burl  \undefined \def \burl#1{\textsf{#1}}\fi
\ifx \doiurl  \undefined \def \doiurl#1{\url{https://doi.org/#1}}\fi
\ifx \betal  \undefined \def \betal{\textit{et al.}}\fi
\ifx \binstitute  \undefined \def \binstitute#1{#1}\fi
\ifx \binstitutionaled  \undefined \def \binstitutionaled#1{#1}\fi
\ifx \bctitle  \undefined \def \bctitle#1{#1}\fi
\ifx \beditor  \undefined \def \beditor#1{#1}\fi
\ifx \bpublisher  \undefined \def \bpublisher#1{#1}\fi
\ifx \bbtitle  \undefined \def \bbtitle#1{#1}\fi
\ifx \bedition  \undefined \def \bedition#1{#1}\fi
\ifx \bseriesno  \undefined \def \bseriesno#1{#1}\fi
\ifx \blocation  \undefined \def \blocation#1{#1}\fi
\ifx \bsertitle  \undefined \def \bsertitle#1{#1}\fi
\ifx \bsnm \undefined \def \bsnm#1{#1}\fi
\ifx \bsuffix \undefined \def \bsuffix#1{#1}\fi
\ifx \bparticle \undefined \def \bparticle#1{#1}\fi
\ifx \barticle \undefined \def \barticle#1{#1}\fi
\bibcommenthead
\ifx \bconfdate \undefined \def \bconfdate #1{#1}\fi
\ifx \botherref \undefined \def \botherref #1{#1}\fi
\ifx \url \undefined \def \url#1{\textsf{#1}}\fi
\ifx \bchapter \undefined \def \bchapter#1{#1}\fi
\ifx \bbook \undefined \def \bbook#1{#1}\fi
\ifx \bcomment \undefined \def \bcomment#1{#1}\fi
\ifx \oauthor \undefined \def \oauthor#1{#1}\fi
\ifx \citeauthoryear \undefined \def \citeauthoryear#1{#1}\fi
\ifx \endbibitem  \undefined \def \endbibitem {}\fi
\ifx \bconflocation  \undefined \def \bconflocation#1{#1}\fi
\ifx \arxivurl  \undefined \def \arxivurl#1{\textsf{#1}}\fi
\csname PreBibitemsHook\endcsname

\bibitem[\protect\citeauthoryear{Zaera}{2014}]{C3CS60374A}
\begin{barticle}
\bauthor{\bsnm{Zaera}, \binits{F.}}:
\batitle{New advances in the use of infrared absorption spectroscopy for the characterization of heterogeneous catalytic reactions}.
\bjtitle{Chem. Soc. Rev.}
\bvolume{43},
\bfpage{7624}--\blpage{7663}
(\byear{2014})
\doiurl{10.1039/C3CS60374A}
\end{barticle}
\endbibitem

\bibitem[\protect\citeauthoryear{Khan et~al.}{2018}]{Khan2018}
\begin{bbook}
\bauthor{\bsnm{Khan}, \binits{S.A.}},
\bauthor{\bsnm{Khan}, \binits{S.B.}},
\bauthor{\bsnm{Khan}, \binits{L.U.}},
\bauthor{\bsnm{Farooq}, \binits{A.}},
\bauthor{\bsnm{Akhtar}, \binits{K.}},
\bauthor{\bsnm{Asiri}, \binits{A.M.}}:
In: \beditor{\bsnm{Sharma}, \binits{S.K.}} (ed.)
\bbtitle{Fourier Transform Infrared Spectroscopy: Fundamentals and Application in Functional Groups and Nanomaterials Characterization},
pp. \bfpage{317}--\blpage{344}.
\bpublisher{Springer},
\blocation{Cham}
(\byear{2018}).
\doiurl{10.1007/978-3-319-92955-2_9}
\end{bbook}
\endbibitem

\bibitem[\protect\citeauthoryear{Rijs and Oomens}{2015}]{Rijs2015}
\begin{bbook}
\bauthor{\bsnm{Rijs}, \binits{A.M.}},
\bauthor{\bsnm{Oomens}, \binits{J.}}:
In: \beditor{\bsnm{Rijs}, \binits{A.M.}},
\beditor{\bsnm{Oomens}, \binits{J.}} (eds.)
\bbtitle{IR Spectroscopic Techniques to Study Isolated Biomolecules},
pp. \bfpage{1}--\blpage{42}.
\bpublisher{Springer},
\blocation{Cham}
(\byear{2015}).
\doiurl{10.1007/128_2014_621}
\end{bbook}
\endbibitem

\bibitem[\protect\citeauthoryear{Costa et~al.}{2015}]{COSTA2015449}
\begin{barticle}
\bauthor{\bsnm{Costa}, \binits{D.}},
\bauthor{\bsnm{Pradier}, \binits{C.-M.}},
\bauthor{\bsnm{Tielens}, \binits{F.}},
\bauthor{\bsnm{Savio}, \binits{L.}}:
\batitle{Adsorption and self-assembly of bio-organic molecules at model surfaces: A route towards increased complexity}.
\bjtitle{Surface Science Reports}
\bvolume{70}(\bissue{4}),
\bfpage{449}--\blpage{553}
(\byear{2015})
\doiurl{10.1016/j.surfrep.2015.10.002}
\end{barticle}
\endbibitem

\bibitem[\protect\citeauthoryear{Hamamoto et~al.}{2015}]{Hamamoto2015}
\begin{barticle}
\bauthor{\bsnm{Hamamoto}, \binits{M.}},
\bauthor{\bsnm{Katsura}, \binits{M.}},
\bauthor{\bsnm{Nishiyama}, \binits{N.}},
\bauthor{\bsnm{Tononue}, \binits{R.}},
\bauthor{\bsnm{Nakashima}, \binits{S.}}:
\batitle{Transmission ir micro-spectroscopy of interfacial water between colloidal silica particles}.
\bjtitle{e-Journal of Surface Science and Nanotechnology}
\bvolume{13},
\bfpage{301}--\blpage{306}
(\byear{2015})
\doiurl{10.1380/ejssnt.2015.301}
\end{barticle}
\endbibitem

\bibitem[\protect\citeauthoryear{Tang et~al.}{2016}]{Tang2016}
\begin{barticle}
\bauthor{\bsnm{Tang}, \binits{M.}},
\bauthor{\bsnm{Cziczo}, \binits{D.J.}},
\bauthor{\bsnm{Grassian}, \binits{V.H.}}:
\batitle{Interactions of water with mineral dust aerosol: Water adsorption, hygroscopicity, cloud condensation, and ice nucleation}.
\bjtitle{Chemical Reviews}
\bvolume{116}(\bissue{7}),
\bfpage{4205}--\blpage{4259}
(\byear{2016})
\doiurl{10.1021/acs.chemrev.5b00529}
\end{barticle}
\endbibitem

\bibitem[\protect\citeauthoryear{Griffith et~al.}{2013}]{Griffith}
\begin{barticle}
\bauthor{\bsnm{Griffith}, \binits{E.C.}},
\bauthor{\bsnm{Guizado}, \binits{T.R.C.}},
\bauthor{\bsnm{Pimentel}, \binits{A.S.}},
\bauthor{\bsnm{Tyndall}, \binits{G.S.}},
\bauthor{\bsnm{Vaida}, \binits{V.}}:
\batitle{Oxidized aromatic–aliphatic mixed films at the air–aqueous solution interface}.
\bjtitle{The Journal of Physical Chemistry C}
\bvolume{117}(\bissue{43}),
\bfpage{22341}--\blpage{22350}
(\byear{2013})
\doiurl{10.1021/jp402737n}
\end{barticle}
\endbibitem

\bibitem[\protect\citeauthoryear{Su et~al.}{2021}]{su_situ_2021}
\begin{barticle}
\bauthor{\bsnm{Su}, \binits{H.}},
\bauthor{\bsnm{Zhou}, \binits{W.}},
\bauthor{\bsnm{Zhou}, \binits{W.}},
\bauthor{\bsnm{Li}, \binits{Y.}},
\bauthor{\bsnm{Zheng}, \binits{L.}},
\bauthor{\bsnm{Zhang}, \binits{H.}},
\bauthor{\bsnm{Liu}, \binits{M.}},
\bauthor{\bsnm{Zhang}, \binits{X.}},
\bauthor{\bsnm{Sun}, \binits{X.}},
\bauthor{\bsnm{Xu}, \binits{Y.}},
\bauthor{\bsnm{Hu}, \binits{F.}},
\bauthor{\bsnm{Zhang}, \binits{J.}},
\bauthor{\bsnm{Hu}, \binits{T.}},
\bauthor{\bsnm{Liu}, \binits{Q.}},
\bauthor{\bsnm{Wei}, \binits{S.}}:
\batitle{In-situ spectroscopic observation of dynamic-coupling oxygen on atomically dispersed iridium electrocatalyst for acidic water oxidation}.
\bjtitle{Nature Communications}
\bvolume{12}(\bissue{1}),
\bfpage{6118}
(\byear{2021})
\doiurl{10.1038/s41467-021-26416-3}
\end{barticle}
\endbibitem

\bibitem[\protect\citeauthoryear{Lai et~al.}{2022}]{Cat_IR_2022}
\begin{barticle}
\bauthor{\bsnm{Lai}, \binits{W.}},
\bauthor{\bsnm{Ma}, \binits{Z.}},
\bauthor{\bsnm{Zhang}, \binits{J.}},
\bauthor{\bsnm{Yuan}, \binits{Y.}},
\bauthor{\bsnm{Qiao}, \binits{Y.}},
\bauthor{\bsnm{Huang}, \binits{H.}}:
\batitle{Dynamic evolution of active sites in electrocatalytic co2 reduction reaction: Fundamental understanding and recent progress}.
\bjtitle{Advanced Functional Materials}
\bvolume{32}(\bissue{16}),
\bfpage{2111193}
(\byear{2022})
\doiurl{10.1002/adfm.202111193}
\end{barticle}
\endbibitem

\bibitem[\protect\citeauthoryear{Deglmann et~al.}{2015}]{Peter2015}
\begin{barticle}
\bauthor{\bsnm{Deglmann}, \binits{P.}},
\bauthor{\bsnm{Sch{\"a}fer}, \binits{A.}},
\bauthor{\bsnm{Lennartz}, \binits{C.}}:
\batitle{Application of quantum calculations in the chemical industry—an overview}.
\bjtitle{International Journal of Quantum Chemistry}
\bvolume{115}(\bissue{3}),
\bfpage{107}--\blpage{136}
(\byear{2015})
\doiurl{10.1002/qua.24811}
\end{barticle}
\endbibitem

\bibitem[\protect\citeauthoryear{Lansford and Vlachos}{2020}]{lansford_infrared_2020}
\begin{barticle}
\bauthor{\bsnm{Lansford}, \binits{J.L.}},
\bauthor{\bsnm{Vlachos}, \binits{D.G.}}:
\batitle{Infrared spectroscopy data- and physics-driven machine learning for characterizing surface microstructure of complex materials}.
\bjtitle{Nature Communications}
\bvolume{11}(\bissue{1}),
\bfpage{1513}
(\byear{2020})
\doiurl{10.1038/s41467-020-15340-7}
\end{barticle}
\endbibitem

\bibitem[\protect\citeauthoryear{Puzzarini et~al.}{2019}]{Interpret_exp_IR}
\begin{barticle}
\bauthor{\bsnm{Puzzarini}, \binits{C.}},
\bauthor{\bsnm{Bloino}, \binits{J.}},
\bauthor{\bsnm{Tasinato}, \binits{N.}},
\bauthor{\bsnm{Barone}, \binits{V.}}:
\batitle{Accuracy and interpretability: The devil and the holy grail. new routes across old boundaries in computational spectroscopy}.
\bjtitle{Chemical Reviews}
\bvolume{119}(\bissue{13}),
\bfpage{8131}--\blpage{8191}
(\byear{2019})
\doiurl{10.1021/acs.chemrev.9b00007}
\end{barticle}
\endbibitem

\bibitem[\protect\citeauthoryear{Larkin}{2011}]{Intepret_exp_IR_2}
\begin{bbook}
\bauthor{\bsnm{Larkin}, \binits{P.}}:
\bbtitle{Infrared and Raman Spectroscopy; Principles and Spectral Interpretation},
(\byear{2011}).
\doiurl{10.1016/C2010-0-68479-3}
\end{bbook}
\endbibitem

\bibitem[\protect\citeauthoryear{Gaigeot and Sprik}{2003}]{Gaigeot2003}
\begin{barticle}
\bauthor{\bsnm{Gaigeot}, \binits{M.-P.}},
\bauthor{\bsnm{Sprik}, \binits{M.}}:
\batitle{Ab initio molecular dynamics computation of the infrared spectrum of aqueous uracil}.
\bjtitle{The Journal of Physical Chemistry B}
\bvolume{107}(\bissue{38}),
\bfpage{10344}--\blpage{10358}
(\byear{2003})
\doiurl{10.1021/jp034788u}
\end{barticle}
\endbibitem

\bibitem[\protect\citeauthoryear{Rodriguez-Betancourtt et~al.}{2011}]{RODRIGUEZBETANCOURTT20111}
\begin{barticle}
\bauthor{\bsnm{Rodriguez-Betancourtt}, \binits{V.-M.}},
\bauthor{\bsnm{Quezada-Navarro}, \binits{V.-M.}},
\bauthor{\bsnm{Neff}, \binits{M.}},
\bauthor{\bsnm{Rauhut}, \binits{G.}}:
\batitle{Anharmonic frequencies of [f,c,n,x] isomers (x=o,s) obtained from explicitly correlated coupled-cluster calculations}.
\bjtitle{Chemical Physics}
\bvolume{387}(\bissue{1}),
\bfpage{1}--\blpage{4}
(\byear{2011})
\doiurl{10.1016/j.chemphys.2011.06.015}
\end{barticle}
\endbibitem

\bibitem[\protect\citeauthoryear{Weymuth et~al.}{2012}]{Weymuth2012MOVIPACVS}
\begin{barticle}
\bauthor{\bsnm{Weymuth}, \binits{T.}},
\bauthor{\bsnm{Haag}, \binits{M.P.}},
\bauthor{\bsnm{Kiewisch}, \binits{K.}},
\bauthor{\bsnm{Luber}, \binits{S.}},
\bauthor{\bsnm{Schenk}, \binits{S.}},
\bauthor{\bsnm{Jacob}, \binits{C.R.}},
\bauthor{\bsnm{Herrmann}, \binits{C.}},
\bauthor{\bsnm{Neugebauer}, \binits{J.}},
\bauthor{\bsnm{Reiher}, \binits{M.}}:
\batitle{Movipac: Vibrational spectroscopy with a robust meta-program for massively parallel standard and inverse calculations}.
\bjtitle{Journal of Computational Chemistry}
\bvolume{33}(\bissue{27}),
\bfpage{2186}--\blpage{2198}
(\byear{2012})
\doiurl{10.1002/jcc.23036}
\end{barticle}
\endbibitem

\bibitem[\protect\citeauthoryear{Thomas et~al.}{2013}]{AIMD_paper}
\begin{barticle}
\bauthor{\bsnm{Thomas}, \binits{M.}},
\bauthor{\bsnm{Brehm}, \binits{M.}},
\bauthor{\bsnm{Fligg}, \binits{R.}},
\bauthor{\bsnm{Vöhringer}, \binits{P.}},
\bauthor{\bsnm{Kirchner}, \binits{B.}}:
\batitle{Computing vibrational spectra from ab initio molecular dynamics}.
\bjtitle{Phys. Chem. Chem. Phys.}
\bvolume{15},
\bfpage{6608}--\blpage{6622}
(\byear{2013})
\doiurl{10.1039/C3CP44302G}
\end{barticle}
\endbibitem

\bibitem[\protect\citeauthoryear{Gaigeot and Spezia}{2015}]{Gaigeot2015}
\begin{bbook}
\bauthor{\bsnm{Gaigeot}, \binits{M.-P.}},
\bauthor{\bsnm{Spezia}, \binits{R.}}:
In: \beditor{\bsnm{Rijs}, \binits{A.M.}},
\beditor{\bsnm{Oomens}, \binits{J.}} (eds.)
\bbtitle{Theoretical Methods for Vibrational Spectroscopy and Collision Induced Dissociation in the Gas Phase},
pp. \bfpage{99}--\blpage{151}.
\bpublisher{Springer},
\blocation{Cham}
(\byear{2015}).
\doiurl{10.1007/128_2014_620}
\end{bbook}
\endbibitem

\bibitem[\protect\citeauthoryear{Miotto and Monacelli}{2024}]{miotto_fast_2024}
\begin{barticle}
\bauthor{\bsnm{Miotto}, \binits{M.}},
\bauthor{\bsnm{Monacelli}, \binits{L.}}:
\batitle{Fast prediction of anharmonic vibrational spectra for complex organic molecules}.
\bjtitle{npj Computational Materials}
\bvolume{10}(\bissue{1}),
\bfpage{1}--\blpage{9}
(\byear{2024})
\doiurl{10.1038/s41524-024-01400-9}
\end{barticle}
\endbibitem

\bibitem[\protect\citeauthoryear{Bastonero and Marzari}{2024}]{bastonero_automated_2024}
\begin{barticle}
\bauthor{\bsnm{Bastonero}, \binits{L.}},
\bauthor{\bsnm{Marzari}, \binits{N.}}:
\batitle{Automated all-functionals infrared and {Raman} spectra}.
\bjtitle{npj Computational Materials}
\bvolume{10}(\bissue{1}),
\bfpage{1}--\blpage{12}
(\byear{2024})
\doiurl{10.1038/s41524-024-01236-3}
\end{barticle}
\endbibitem

\bibitem[\protect\citeauthoryear{Behler and Parrinello}{2007}]{Behler_2007}
\begin{barticle}
\bauthor{\bsnm{Behler}, \binits{J.}},
\bauthor{\bsnm{Parrinello}, \binits{M.}}:
\batitle{Generalized neural-network representation of high-dimensional potential-energy surfaces}.
\bjtitle{Phys. Rev. Lett.}
\bvolume{98},
\bfpage{146401}
(\byear{2007})
\doiurl{10.1103/PhysRevLett.98.146401}
\end{barticle}
\endbibitem

\bibitem[\protect\citeauthoryear{Bartók}{2010}]{bartok_gaussian_2010}
\begin{botherref}
\oauthor{\bsnm{Bartók}, \binits{A.P.}}:
Gaussian {Approximation} {Potentials}: {The} {Accuracy} of {Quantum} {Mechanics}, without the {Electrons}.
Physical Review Letters
\textbf{104}(13)
(2010)
\doiurl{10.1103/PhysRevLett.104.136403}
\end{botherref}
\endbibitem

\bibitem[\protect\citeauthoryear{Smith et~al.}{2017}]{smith_ani-1_2017}
\begin{barticle}
\bauthor{\bsnm{Smith}, \binits{J.S.}},
\bauthor{\bsnm{Isayev}, \binits{O.}},
\bauthor{\bsnm{Roitberg}, \binits{A.E.}}:
\batitle{{ANI}-1: an extensible neural network potential with {DFT} accuracy at force field computational cost}.
\bjtitle{Chemical Science}
\bvolume{8}(\bissue{4}),
\bfpage{3192}--\blpage{3203}
(\byear{2017})
\doiurl{10.1039/C6SC05720A}
\end{barticle}
\endbibitem

\bibitem[\protect\citeauthoryear{Gastegger et~al.}{2017}]{gastegger_machine_2017}
\begin{barticle}
\bauthor{\bsnm{Gastegger}, \binits{M.}},
\bauthor{\bsnm{Behler}, \binits{J.}},
\bauthor{\bsnm{Marquetand}, \binits{P.}}:
\batitle{Machine learning molecular dynamics for the simulation of infrared spectra}.
\bjtitle{Chemical Science}
\bvolume{8}(\bissue{10}),
\bfpage{6924}--\blpage{6935}
(\byear{2017})
\doiurl{10.1039/C7SC02267K}
\end{barticle}
\endbibitem

\bibitem[\protect\citeauthoryear{Schütt et~al.}{2018}]{SchNet}
\begin{barticle}
\bauthor{\bsnm{Schütt}, \binits{K.T.}},
\bauthor{\bsnm{Sauceda}, \binits{H.E.}},
\bauthor{\bsnm{Kindermans}, \binits{P.-J.}},
\bauthor{\bsnm{Tkatchenko}, \binits{A.}},
\bauthor{\bsnm{Müller}, \binits{K.-R.}}:
\batitle{{SchNet – A deep learning architecture for molecules and materials}}.
\bjtitle{The Journal of Chemical Physics}
\bvolume{148}(\bissue{24}),
\bfpage{241722}
(\byear{2018})
\doiurl{10.1063/1.5019779}
\end{barticle}
\endbibitem

\bibitem[\protect\citeauthoryear{Zaverkin and Kästner}{2020}]{GM_2020}
\begin{barticle}
\bauthor{\bsnm{Zaverkin}, \binits{V.}},
\bauthor{\bsnm{Kästner}, \binits{J.}}:
\batitle{Gaussian moments as physically inspired molecular descriptors for accurate and scalable machine learning potentials}.
\bjtitle{Journal of Chemical Theory and Computation}
\bvolume{16}(\bissue{8}),
\bfpage{5410}--\blpage{5421}
(\byear{2020})
\doiurl{10.1021/acs.jctc.0c00347}
\end{barticle}
\endbibitem

\bibitem[\protect\citeauthoryear{Zaverkin et~al.}{2021}]{GM_2021}
\begin{barticle}
\bauthor{\bsnm{Zaverkin}, \binits{V.}},
\bauthor{\bsnm{Holzmüller}, \binits{D.}},
\bauthor{\bsnm{Steinwart}, \binits{I.}},
\bauthor{\bsnm{Kästner}, \binits{J.}}:
\batitle{Fast and sample-efficient interatomic neural network potentials for molecules and materials based on gaussian moments}.
\bjtitle{Journal of Chemical Theory and Computation}
\bvolume{17}(\bissue{10}),
\bfpage{6658}--\blpage{6670}
(\byear{2021})
\doiurl{10.1021/acs.jctc.1c00527}
\end{barticle}
\endbibitem

\bibitem[\protect\citeauthoryear{Deringer et~al.}{2021}]{GPR}
\begin{barticle}
\bauthor{\bsnm{Deringer}, \binits{V.L.}},
\bauthor{\bsnm{Bartók}, \binits{A.P.}},
\bauthor{\bsnm{Bernstein}, \binits{N.}},
\bauthor{\bsnm{Wilkins}, \binits{D.M.}},
\bauthor{\bsnm{Ceriotti}, \binits{M.}},
\bauthor{\bsnm{Csányi}, \binits{G.}}:
\batitle{Gaussian process regression for materials and molecules}.
\bjtitle{Chemical Reviews}
\bvolume{121}(\bissue{16}),
\bfpage{10073}--\blpage{10141}
(\byear{2021})
\doiurl{10.1021/acs.chemrev.1c00022}
\end{barticle}
\endbibitem

\bibitem[\protect\citeauthoryear{Batzner et~al.}{2022}]{NequiP}
\begin{barticle}
\bauthor{\bsnm{Batzner}, \binits{S.}},
\bauthor{\bsnm{Musaelian}, \binits{A.}},
\bauthor{\bsnm{Sun}, \binits{L.}},
\bauthor{\bsnm{Geiger}, \binits{M.}},
\bauthor{\bsnm{Mailoa}, \binits{J.P.}},
\bauthor{\bsnm{Kornbluth}, \binits{M.}},
\bauthor{\bsnm{Molinari}, \binits{N.}},
\bauthor{\bsnm{Smidt}, \binits{T.E.}},
\bauthor{\bsnm{Kozinsky}, \binits{B.}}:
\batitle{E(3)-equivariant graph neural networks for data-efficient and accurate interatomic potentials}.
\bjtitle{Nature Communications}
\bvolume{13}(\bissue{1}),
\bfpage{2453}
(\byear{2022})
\doiurl{10.1038/s41467-022-29939-5}
\end{barticle}
\endbibitem

\bibitem[\protect\citeauthoryear{Musaelian et~al.}{2023}]{Allegro}
\begin{barticle}
\bauthor{\bsnm{Musaelian}, \binits{A.}},
\bauthor{\bsnm{Batzner}, \binits{S.}},
\bauthor{\bsnm{Johansson}, \binits{A.}},
\bauthor{\bsnm{Sun}, \binits{L.}},
\bauthor{\bsnm{Owen}, \binits{C.J.}},
\bauthor{\bsnm{Kornbluth}, \binits{M.}},
\bauthor{\bsnm{Kozinsky}, \binits{B.}}:
\batitle{Learning local equivariant representations for large-scale atomistic dynamics}.
\bjtitle{Nature Communications}
\bvolume{14}(\bissue{1}),
\bfpage{579}
(\byear{2023})
\doiurl{10.1038/s41467-023-36329-y}
\end{barticle}
\endbibitem

\bibitem[\protect\citeauthoryear{Grisafi et~al.}{2018}]{grisafi_symmetry-adapted_2018}
\begin{barticle}
\bauthor{\bsnm{Grisafi}, \binits{A.}},
\bauthor{\bsnm{Wilkins}, \binits{D.M.}},
\bauthor{\bsnm{Csányi}, \binits{G.}},
\bauthor{\bsnm{Ceriotti}, \binits{M.}}:
\batitle{Symmetry-{Adapted} {Machine} {Learning} for {Tensorial} {Properties} of {Atomistic} {Systems}}.
\bjtitle{Physical Review Letters}
\bvolume{120}(\bissue{3}),
\bfpage{036002}
(\byear{2018})
\doiurl{10.1103/PhysRevLett.120.036002}
\end{barticle}
\endbibitem

\bibitem[\protect\citeauthoryear{Unke and Meuwly}{2019}]{PhysNet}
\begin{barticle}
\bauthor{\bsnm{Unke}, \binits{O.T.}},
\bauthor{\bsnm{Meuwly}, \binits{M.}}:
\batitle{Physnet: A neural network for predicting energies, forces, dipole moments, and partial charges}.
\bjtitle{Journal of Chemical Theory and Computation}
\bvolume{15}(\bissue{6}),
\bfpage{3678}--\blpage{3693}
(\byear{2019})
\doiurl{10.1021/acs.jctc.9b00181}
\end{barticle}
\endbibitem

\bibitem[\protect\citeauthoryear{Gastegger et~al.}{2021}]{gastegger_machine_2021}
\begin{barticle}
\bauthor{\bsnm{Gastegger}, \binits{M.}},
\bauthor{\bsnm{Schütt}, \binits{K.T.}},
\bauthor{\bsnm{Müller}, \binits{K.-R.}}:
\batitle{Machine learning of solvent effects on molecular spectra and reactions}.
\bjtitle{Chemical Science}
\bvolume{12}(\bissue{34}),
\bfpage{11473}--\blpage{11483}
(\byear{2021})
\doiurl{10.1039/D1SC02742E}
\end{barticle}
\endbibitem

\bibitem[\protect\citeauthoryear{Sch{\"u}tt et~al.}{2021}]{PaiNN}
\begin{bchapter}
\bauthor{\bsnm{Sch{\"u}tt}, \binits{K.}},
\bauthor{\bsnm{Unke}, \binits{O.}},
\bauthor{\bsnm{Gastegger}, \binits{M.}}:
\bctitle{Equivariant message passing for the prediction of tensorial properties and molecular spectra}.
In: \beditor{\bsnm{Meila}, \binits{M.}},
\beditor{\bsnm{Zhang}, \binits{T.}} (eds.)
\bbtitle{Proceedings of the 38th International Conference on Machine Learning}.
\bsertitle{Proceedings of Machine Learning Research},
vol. \bseriesno{139},
pp. \bfpage{9377}--\blpage{9388}.
\bpublisher{PMLR}, \blocation{???}
(\byear{2021}).
\burl{https://proceedings.mlr.press/v139/schutt21a.html}
\end{bchapter}
\endbibitem

\bibitem[\protect\citeauthoryear{Beckmann et~al.}{2022}]{beckmann_infrared_2022}
\begin{barticle}
\bauthor{\bsnm{Beckmann}, \binits{R.}},
\bauthor{\bsnm{Brieuc}, \binits{F.}},
\bauthor{\bsnm{Schran}, \binits{C.}},
\bauthor{\bsnm{Marx}, \binits{D.}}:
\batitle{Infrared {Spectra} at {Coupled} {Cluster} {Accuracy} from {Neural} {Network} {Representations}}.
\bjtitle{Journal of Chemical Theory and Computation}
\bvolume{18}(\bissue{9}),
\bfpage{5492}--\blpage{5501}
(\byear{2022})
\doiurl{10.1021/acs.jctc.2c00511}
\end{barticle}
\endbibitem

\bibitem[\protect\citeauthoryear{Batatia et~al.}{2022a}]{Batatia2022Design}
\begin{botherref}
\oauthor{\bsnm{Batatia}, \binits{I.}},
\oauthor{\bsnm{Batzner}, \binits{S.}},
\oauthor{\bsnm{Kov{\'a}cs}, \binits{D.P.}},
\oauthor{\bsnm{Musaelian}, \binits{A.}},
\oauthor{\bsnm{Simm}, \binits{G.N.C.}},
\oauthor{\bsnm{Drautz}, \binits{R.}},
\oauthor{\bsnm{Ortner}, \binits{C.}},
\oauthor{\bsnm{Kozinsky}, \binits{B.}},
\oauthor{\bsnm{Cs{\'a}nyi}, \binits{G.}}:
The Design Space of E(3)-Equivariant Atom-Centered Interatomic Potentials
(2022).
\doiurl{10.48550/arXiv.2205.06643}
\end{botherref}
\endbibitem

\bibitem[\protect\citeauthoryear{Batatia et~al.}{2022b}]{MACE}
\begin{barticle}
\bauthor{\bsnm{Batatia}, \binits{I.}},
\bauthor{\bsnm{Kovacs}, \binits{D.P.}},
\bauthor{\bsnm{Simm}, \binits{G.}},
\bauthor{\bsnm{Ortner}, \binits{C.}},
\bauthor{\bsnm{Cs{\'a}nyi}, \binits{G.}}:
\batitle{Mace: Higher order equivariant message passing neural networks for fast and accurate force fields}.
\bjtitle{Advances in Neural Information Processing Systems}
\bvolume{35},
\bfpage{11423}--\blpage{11436}
(\byear{2022})
\end{barticle}
\endbibitem

\bibitem[\protect\citeauthoryear{Tang et~al.}{2023}]{tang_machine_2023}
\begin{barticle}
\bauthor{\bsnm{Tang}, \binits{Z.}},
\bauthor{\bsnm{Bromley}, \binits{S.T.}},
\bauthor{\bsnm{Hammer}, \binits{B.}}:
\batitle{A machine learning potential for simulating infrared spectra of nanosilicate clusters}.
\bjtitle{The Journal of Chemical Physics}
\bvolume{158}(\bissue{22}),
\bfpage{224108}
(\byear{2023})
\doiurl{10.1063/5.0150379}
\end{barticle}
\endbibitem

\bibitem[\protect\citeauthoryear{Zou et~al.}{2023}]{zou_deep_2023}
\begin{barticle}
\bauthor{\bsnm{Zou}, \binits{Z.}},
\bauthor{\bsnm{Zhang}, \binits{Y.}},
\bauthor{\bsnm{Liang}, \binits{L.}},
\bauthor{\bsnm{Wei}, \binits{M.}},
\bauthor{\bsnm{Leng}, \binits{J.}},
\bauthor{\bsnm{Jiang}, \binits{J.}},
\bauthor{\bsnm{Luo}, \binits{Y.}},
\bauthor{\bsnm{Hu}, \binits{W.}}:
\batitle{A deep learning model for predicting selected organic molecular spectra}.
\bjtitle{Nature Computational Science}
\bvolume{3}(\bissue{11}),
\bfpage{957}--\blpage{964}
(\byear{2023})
\doiurl{10.1038/s43588-023-00550-y}
\end{barticle}
\endbibitem

\bibitem[\protect\citeauthoryear{Stienstra et~al.}{2024}]{stienstra_graphormer-ir_2024}
\begin{barticle}
\bauthor{\bsnm{Stienstra}, \binits{C.M.K.}},
\bauthor{\bsnm{Hebert}, \binits{L.}},
\bauthor{\bsnm{Thomas}, \binits{P.}},
\bauthor{\bsnm{Haack}, \binits{A.}},
\bauthor{\bsnm{Guo}, \binits{J.}},
\bauthor{\bsnm{Hopkins}, \binits{W.S.}}:
\batitle{Graphormer-{IR}: {Graph} {Transformers} {Predict} {Experimental} {IR} {Spectra} {Using} {Highly} {Specialized} {Attention}}.
\bjtitle{Journal of Chemical Information and Modeling}
\bvolume{64}(\bissue{12}),
\bfpage{4613}--\blpage{4629}
(\byear{2024})
\doiurl{10.1021/acs.jcim.4c00378}
\end{barticle}
\endbibitem

\bibitem[\protect\citeauthoryear{Krzyżanowski and Matyszczak}{2024}]{krzyzanowski_machine_2024}
\begin{barticle}
\bauthor{\bsnm{Krzyżanowski}, \binits{M.}},
\bauthor{\bsnm{Matyszczak}, \binits{G.}}:
\batitle{Machine learning prediction of organic moieties from the {IR} spectra, enhanced by additionally using the derivative {IR} data}.
\bjtitle{Chemical Papers}
\bvolume{78}(\bissue{5}),
\bfpage{3149}--\blpage{3173}
(\byear{2024})
\doiurl{10.1007/s11696-024-03301-z}
\end{barticle}
\endbibitem

\bibitem[\protect\citeauthoryear{Yuan et~al.}{2025}]{yuan2025qme14scomprehensiveefficientspectral}
\begin{botherref}
\oauthor{\bsnm{Yuan}, \binits{M.}},
\oauthor{\bsnm{Zou}, \binits{Z.}},
\oauthor{\bsnm{Hu}, \binits{W.}}:
QMe14S, A Comprehensive and Efficient Spectral Dataset for Small Organic Molecules
(2025).
\url{https://arxiv.org/abs/2501.18876}
\end{botherref}
\endbibitem

\bibitem[\protect\citeauthoryear{Rowe et~al.}{2020}]{General_MLIP_1}
\begin{barticle}
\bauthor{\bsnm{Rowe}, \binits{P.}},
\bauthor{\bsnm{Deringer}, \binits{V.L.}},
\bauthor{\bsnm{Gasparotto}, \binits{P.}},
\bauthor{\bsnm{Csányi}, \binits{G.}},
\bauthor{\bsnm{Michaelides}, \binits{A.}}:
\batitle{{An accurate and transferable machine learning potential for carbon}}.
\bjtitle{The Journal of Chemical Physics}
\bvolume{153}(\bissue{3}),
\bfpage{034702}
(\byear{2020})
\doiurl{10.1063/5.0005084}
\end{barticle}
\endbibitem

\bibitem[\protect\citeauthoryear{Bart{\'o}k et~al.}{2018}]{bartok2018machine}
\begin{barticle}
\bauthor{\bsnm{Bart{\'o}k}, \binits{A.P.}},
\bauthor{\bsnm{Kermode}, \binits{J.}},
\bauthor{\bsnm{Bernstein}, \binits{N.}},
\bauthor{\bsnm{Cs{\'a}nyi}, \binits{G.}}:
\batitle{Machine learning a general-purpose interatomic potential for silicon}.
\bjtitle{Physical Review X}
\bvolume{8}(\bissue{4}),
\bfpage{041048}
(\byear{2018})
\end{barticle}
\endbibitem

\bibitem[\protect\citeauthoryear{Deringer et~al.}{2020}]{deringer_general-purpose_2020}
\begin{barticle}
\bauthor{\bsnm{Deringer}, \binits{V.L.}},
\bauthor{\bsnm{Caro}, \binits{M.A.}},
\bauthor{\bsnm{Csányi}, \binits{G.}}:
\batitle{A general-purpose machine-learning force field for bulk and nanostructured phosphorus}.
\bjtitle{Nature Communications}
\bvolume{11}(\bissue{1}),
\bfpage{5461}
(\byear{2020})
\doiurl{10.1038/s41467-020-19168-z}
\end{barticle}
\endbibitem

\bibitem[\protect\citeauthoryear{Podryabinkin and Shapeev}{2017}]{AL_1}
\begin{barticle}
\bauthor{\bsnm{Podryabinkin}, \binits{E.V.}},
\bauthor{\bsnm{Shapeev}, \binits{A.V.}}:
\batitle{Active learning of linearly parametrized interatomic potentials}.
\bjtitle{Computational Materials Science}
\bvolume{140},
\bfpage{171}--\blpage{180}
(\byear{2017})
\doiurl{10.1016/j.commatsci.2017.08.031}
\end{barticle}
\endbibitem

\bibitem[\protect\citeauthoryear{Gubaev et~al.}{2019}]{AL_2}
\begin{barticle}
\bauthor{\bsnm{Gubaev}, \binits{K.}},
\bauthor{\bsnm{Podryabinkin}, \binits{E.V.}},
\bauthor{\bsnm{Hart}, \binits{G.L.W.}},
\bauthor{\bsnm{Shapeev}, \binits{A.V.}}:
\batitle{Accelerating high-throughput searches for new alloys with active learning of interatomic potentials}.
\bjtitle{Computational Materials Science}
\bvolume{156},
\bfpage{148}--\blpage{156}
(\byear{2019})
\doiurl{10.1016/j.commatsci.2018.09.031}
\end{barticle}
\endbibitem

\bibitem[\protect\citeauthoryear{Vandermause et~al.}{2020}]{Flare_1}
\begin{barticle}
\bauthor{\bsnm{Vandermause}, \binits{J.}},
\bauthor{\bsnm{Torrisi}, \binits{S.B.}},
\bauthor{\bsnm{Batzner}, \binits{S.}},
\bauthor{\bsnm{Xie}, \binits{Y.}},
\bauthor{\bsnm{Sun}, \binits{L.}},
\bauthor{\bsnm{Kolpak}, \binits{A.M.}},
\bauthor{\bsnm{Kozinsky}, \binits{B.}}:
\batitle{On-the-fly active learning of interpretable {Bayesian} force fields for atomistic rare events}.
\bjtitle{npj Computational Materials}
\bvolume{6}(\bissue{1}),
\bfpage{1}--\blpage{11}
(\byear{2020})
\doiurl{10.1038/s41524-020-0283-z}
\end{barticle}
\endbibitem

\bibitem[\protect\citeauthoryear{van~der Oord et~al.}{2023}]{van_der_oord_hyperactive_2023}
\begin{barticle}
\bauthor{\bsnm{Oord}, \binits{C.}},
\bauthor{\bsnm{Sachs}, \binits{M.}},
\bauthor{\bsnm{Kovács}, \binits{D.P.}},
\bauthor{\bsnm{Ortner}, \binits{C.}},
\bauthor{\bsnm{Csányi}, \binits{G.}}:
\batitle{Hyperactive learning for data-driven interatomic potentials}.
\bjtitle{npj Computational Materials}
\bvolume{9}(\bissue{1}),
\bfpage{1}--\blpage{14}
(\byear{2023})
\doiurl{10.1038/s41524-023-01104-6}
\end{barticle}
\endbibitem

\bibitem[\protect\citeauthoryear{Kulichenko et~al.}{2023}]{AL_uncertainty-driven_2023}
\begin{barticle}
\bauthor{\bsnm{Kulichenko}, \binits{M.}},
\bauthor{\bsnm{Barros}, \binits{K.}},
\bauthor{\bsnm{Lubbers}, \binits{N.}},
\bauthor{\bsnm{Li}, \binits{Y.W.}},
\bauthor{\bsnm{Messerly}, \binits{R.}},
\bauthor{\bsnm{Tretiak}, \binits{S.}},
\bauthor{\bsnm{Smith}, \binits{J.S.}},
\bauthor{\bsnm{Nebgen}, \binits{B.}}:
\batitle{Uncertainty-driven dynamics for active learning of interatomic potentials}.
\bjtitle{Nature Computational Science}
\bvolume{3}(\bissue{3}),
\bfpage{230}--\blpage{239}
(\byear{2023})
\doiurl{10.1038/s43588-023-00406-5}
\end{barticle}
\endbibitem

\bibitem[\protect\citeauthoryear{Tan et~al.}{2023}]{tan_single-model_2023}
\begin{barticle}
\bauthor{\bsnm{Tan}, \binits{A.R.}},
\bauthor{\bsnm{Urata}, \binits{S.}},
\bauthor{\bsnm{Goldman}, \binits{S.}},
\bauthor{\bsnm{Dietschreit}, \binits{J.C.B.}},
\bauthor{\bsnm{Gómez-Bombarelli}, \binits{R.}}:
\batitle{Single-model uncertainty quantification in neural network potentials does not consistently outperform model ensembles}.
\bjtitle{npj Computational Materials}
\bvolume{9}(\bissue{1}),
\bfpage{1}--\blpage{11}
(\byear{2023})
\doiurl{10.1038/s41524-023-01180-8}
\end{barticle}
\endbibitem

\bibitem[\protect\citeauthoryear{Zaverkin et~al.}{2024}]{zaverkin_uncertainty-biased_2024}
\begin{barticle}
\bauthor{\bsnm{Zaverkin}, \binits{V.}},
\bauthor{\bsnm{Holzmüller}, \binits{D.}},
\bauthor{\bsnm{Christiansen}, \binits{H.}},
\bauthor{\bsnm{Errica}, \binits{F.}},
\bauthor{\bsnm{Alesiani}, \binits{F.}},
\bauthor{\bsnm{Takamoto}, \binits{M.}},
\bauthor{\bsnm{Niepert}, \binits{M.}},
\bauthor{\bsnm{Kästner}, \binits{J.}}:
\batitle{Uncertainty-biased molecular dynamics for learning uniformly accurate interatomic potentials}.
\bjtitle{npj Computational Materials}
\bvolume{10}(\bissue{1}),
\bfpage{1}--\blpage{18}
(\byear{2024})
\doiurl{10.1038/s41524-024-01254-1}
\end{barticle}
\endbibitem

\bibitem[\protect\citeauthoryear{Ghosh et~al.}{2025}]{Ghosh/etal:2025}
\begin{barticle}
\bauthor{\bsnm{Ghosh}, \binits{K.}},
\bauthor{\bsnm{Todorovi{\'{c}}}, \binits{M.}},
\bauthor{\bsnm{Vehtari}, \binits{A.}},
\bauthor{\bsnm{Rinke}, \binits{P.}}:
\batitle{Active learning of molecular data for task-specific objectives}.
\bjtitle{J. Chem. Phys.}
\bvolume{162}(\bissue{1}),
\bfpage{014103}
(\byear{2025})
\end{barticle}
\endbibitem

\bibitem[\protect\citeauthoryear{Homm et~al.}{2025}]{Homm/Laakso/Rinke:2025}
\begin{barticle}
\bauthor{\bsnm{Homm}, \binits{H.}},
\bauthor{\bsnm{Laakso}, \binits{J.}},
\bauthor{\bsnm{Rinke}, \binits{P.}}:
\batitle{Efficient dataset generation for machine learning halide perovskite alloys}.
\bjtitle{Phys. Rev. Mater.}
\bvolume{9},
\bfpage{053802}
(\byear{2025})
\end{barticle}
\endbibitem

\bibitem[\protect\citeauthoryear{Bhatia et~al.}{2024}]{PALIRS_repo}
\begin{botherref}
\oauthor{\bsnm{Bhatia}, \binits{N.}},
\oauthor{\bsnm{Krejčí}, \binits{O.}},
\oauthor{\bsnm{Rinke}, \binits{P.}}:
PALIRS.
GitLab
(2024).
\url{https://gitlab.com/cest-group/PALIRS}
\end{botherref}
\endbibitem

\bibitem[\protect\citeauthoryear{Schran et~al.}{2020}]{schran_committee_2020}
\begin{barticle}
\bauthor{\bsnm{Schran}, \binits{C.}},
\bauthor{\bsnm{Brezina}, \binits{K.}},
\bauthor{\bsnm{Marsalek}, \binits{O.}}:
\batitle{Committee neural network potentials control generalization errors and enable active learning}.
\bjtitle{The Journal of Chemical Physics}
\bvolume{153}(\bissue{10}),
\bfpage{104105}
(\byear{2020})
\doiurl{10.1063/5.0016004}
\end{barticle}
\endbibitem

\bibitem[\protect\citeauthoryear{Smith et~al.}{2018}]{QBC_2}
\begin{barticle}
\bauthor{\bsnm{Smith}, \binits{J.S.}},
\bauthor{\bsnm{Nebgen}, \binits{B.}},
\bauthor{\bsnm{Lubbers}, \binits{N.}},
\bauthor{\bsnm{Isayev}, \binits{O.}},
\bauthor{\bsnm{Roitberg}, \binits{A.E.}}:
\batitle{{Less is more: Sampling chemical space with active learning}}.
\bjtitle{The Journal of Chemical Physics}
\bvolume{148}(\bissue{24}),
\bfpage{241733}
(\byear{2018})
\doiurl{10.1063/1.5023802}
\end{barticle}
\endbibitem

\bibitem[\protect\citeauthoryear{Qu et~al.}{2018}]{qu_permutationally_2018}
\begin{barticle}
\bauthor{\bsnm{Qu}, \binits{C.}},
\bauthor{\bsnm{Yu}, \binits{Q.}},
\bauthor{\bsnm{Bowman}, \binits{J.M.}}:
\batitle{Permutationally {Invariant} {Potential} {Energy} {Surfaces}}.
\bjtitle{Annual Review of Physical Chemistry}
\bvolume{69}(\bissue{Volume 69, 2018}),
\bfpage{151}--\blpage{175}
(\byear{2018})
\doiurl{10.1146/annurev-physchem-050317-021139}
\end{barticle}
\endbibitem

\bibitem[\protect\citeauthoryear{Blum et~al.}{2009}]{blum2009ab}
\begin{barticle}
\bauthor{\bsnm{Blum}, \binits{V.}},
\bauthor{\bsnm{Gehrke}, \binits{R.}},
\bauthor{\bsnm{Hanke}, \binits{F.}},
\bauthor{\bsnm{Havu}, \binits{P.}},
\bauthor{\bsnm{Havu}, \binits{V.}},
\bauthor{\bsnm{Ren}, \binits{X.}},
\bauthor{\bsnm{Reuter}, \binits{K.}},
\bauthor{\bsnm{Scheffler}, \binits{M.}}:
\batitle{Ab initio molecular simulations with numeric atom-centered orbitals}.
\bjtitle{Computer Physics Communications}
\bvolume{180}(\bissue{11}),
\bfpage{2175}--\blpage{2196}
(\byear{2009})
\doiurl{10.1016/j.cpc.2009.06.022}
\end{barticle}
\endbibitem

\bibitem[\protect\citeauthoryear{Havu et~al.}{2009}]{havu2009efficient}
\begin{barticle}
\bauthor{\bsnm{Havu}, \binits{V.}},
\bauthor{\bsnm{Blum}, \binits{V.}},
\bauthor{\bsnm{Havu}, \binits{P.}},
\bauthor{\bsnm{Scheffler}, \binits{M.}}:
\batitle{Efficient {$O(N)$} integration for all-electron electronic structure calculation using numeric basis functions}.
\bjtitle{Journal of Computational Physics}
\bvolume{228}(\bissue{22}),
\bfpage{8367}--\blpage{8379}
(\byear{2009})
\doiurl{10.1016/j.jcp.2009.08.008}
\end{barticle}
\endbibitem

\bibitem[\protect\citeauthoryear{Levchenko et~al.}{2015}]{levchenko2015hybrid}
\begin{barticle}
\bauthor{\bsnm{Levchenko}, \binits{S.V.}},
\bauthor{\bsnm{Ren}, \binits{X.}},
\bauthor{\bsnm{Wieferink}, \binits{J.}},
\bauthor{\bsnm{Johanni}, \binits{R.}},
\bauthor{\bsnm{Rinke}, \binits{P.}},
\bauthor{\bsnm{Blum}, \binits{V.}},
\bauthor{\bsnm{Scheffler}, \binits{M.}}:
\batitle{Hybrid functionals for large periodic systems in an all-electron, numeric atom-centered basis framework}.
\bjtitle{Computer Physics Communications}
\bvolume{192},
\bfpage{60}--\blpage{69}
(\byear{2015})
\doiurl{10.1016/j.cpc.2015.02.021}
\end{barticle}
\endbibitem

\bibitem[\protect\citeauthoryear{Ren et~al.}{2012}]{Xinguo/implem_full_author_list}
\begin{barticle}
\bauthor{\bsnm{Ren}, \binits{X.}},
\bauthor{\bsnm{Rinke}, \binits{P.}},
\bauthor{\bsnm{Blum}, \binits{V.}},
\bauthor{\bsnm{Wieferink}, \binits{J.}},
\bauthor{\bsnm{Tkatchenko}, \binits{A.}},
\bauthor{\bsnm{Andrea}, \binits{S.}},
\bauthor{\bsnm{Reuter}, \binits{K.}},
\bauthor{\bsnm{Blum}, \binits{V.}},
\bauthor{\bsnm{Scheffler}, \binits{M.}}:
\batitle{Resolution-of-identity approach to {Hartree}-{Fock}, hybrid density functionals, {RPA}, {MP2}, and {GW} with numeric atom-centered orbital basis functions}.
\bjtitle{New J. Phys.}
\bvolume{14},
\bfpage{053020}
(\byear{2012})
\end{barticle}
\endbibitem

\bibitem[\protect\citeauthoryear{Huo and Rupp}{2022}]{Huo_2022}
\begin{barticle}
\bauthor{\bsnm{Huo}, \binits{H.}},
\bauthor{\bsnm{Rupp}, \binits{M.}}:
\batitle{Unified representation of molecules and crystals for machine learning}.
\bjtitle{Machine Learning: Science and Technology}
\bvolume{3}(\bissue{4}),
\bfpage{045017}
(\byear{2022})
\doiurl{10.1088/2632-2153/aca005}
\end{barticle}
\endbibitem

\bibitem[\protect\citeauthoryear{Himanen et~al.}{2020}]{himanen2020dscribe}
\begin{barticle}
\bauthor{\bsnm{Himanen}, \binits{L.}},
\bauthor{\bsnm{Jäger}, \binits{M.O.J.}},
\bauthor{\bsnm{Morooka}, \binits{E.V.}},
\bauthor{\bsnm{{Federici Canova}}, \binits{F.}},
\bauthor{\bsnm{Ranawat}, \binits{Y.S.}},
\bauthor{\bsnm{Gao}, \binits{D.Z.}},
\bauthor{\bsnm{Rinke}, \binits{P.}},
\bauthor{\bsnm{Foster}, \binits{A.S.}}:
\batitle{Dscribe: Library of descriptors for machine learning in materials science}.
\bjtitle{Computer Physics Communications}
\bvolume{247},
\bfpage{106949}
(\byear{2020})
\doiurl{10.1016/j.cpc.2019.106949}
\end{barticle}
\endbibitem

\bibitem[\protect\citeauthoryear{Esch et~al.}{2021}]{esch_quantitative_2021}
\begin{barticle}
\bauthor{\bsnm{Esch}, \binits{B.v.d.}},
\bauthor{\bsnm{Peters}, \binits{L.D.M.}},
\bauthor{\bsnm{Sauerland}, \binits{L.}},
\bauthor{\bsnm{Ochsenfeld}, \binits{C.}}:
\batitle{Quantitative {Comparison} of {Experimental} and {Computed} {IR}-{Spectra} {Extracted} from {Ab} {Initio} {Molecular} {Dynamics}}.
\bjtitle{Journal of Chemical Theory and Computation}
\bvolume{17}(\bissue{2}),
\bfpage{985}--\blpage{995}
(\byear{2021})
\doiurl{10.1021/acs.jctc.0c01279}
\end{barticle}
\endbibitem

\bibitem[\protect\citeauthoryear{Wallace}{}]{Wallace2024}
\begin{botherref}
\oauthor{\bsnm{Wallace}, \binits{W.E.}}:
{NIST Chemistry WebBook, NIST Standard Reference Database Number 69}.
Linstrom, P. J. and Mallard, W. G.; National Institute of Standards and Technology; Gaithersburg, MD.
Vol. 20899, (retrieved September 12, 2024); Chapter "Infrared Spectra" by NIST Mass Spectrometry Data Center.
\doiurl{10.18434/T4D303}
\end{botherref}
\endbibitem

\bibitem[\protect\citeauthoryear{Sagiv et~al.}{2018}]{sagiv_anharmonic_2018}
\begin{barticle}
\bauthor{\bsnm{Sagiv}, \binits{L.}},
\bauthor{\bsnm{Hirshberg}, \binits{B.}},
\bauthor{\bsnm{Gerber}, \binits{R.B.}}:
\batitle{Anharmonic vibrational spectroscopy calculations using the \textit{ab initio} {CSP} method: {Applications} to {H2CO3}, ({H2CO3})2, {H2CO3}-{H2O} and isotopologues}.
\bjtitle{Chemical Physics}
\bvolume{514},
\bfpage{44}--\blpage{54}
(\byear{2018})
\doiurl{10.1016/j.chemphys.2017.12.015}
\end{barticle}
\endbibitem

\bibitem[\protect\citeauthoryear{Perdew et~al.}{1997}]{perdew_generalized_1997}
\begin{barticle}
\bauthor{\bsnm{Perdew}, \binits{J.P.}},
\bauthor{\bsnm{Burke}, \binits{K.}},
\bauthor{\bsnm{Ernzerhof}, \binits{M.}}:
\batitle{Generalized {Gradient} {Approximation} {Made} {Simple} [{Phys}. {Rev}. {Lett}. 77, 3865 (1996)]}.
\bjtitle{Physical Review Letters}
\bvolume{78}(\bissue{7}),
\bfpage{1396}--\blpage{1396}
(\byear{1997})
\doiurl{10.1103/PhysRevLett.78.1396}
\end{barticle}
\endbibitem

\bibitem[\protect\citeauthoryear{Lenthe et~al.}{1993}]{lenthe1993relativistic}
\begin{barticle}
\bauthor{\bsnm{Lenthe}, \binits{E.v.}},
\bauthor{\bsnm{Baerends}, \binits{E.J.}},
\bauthor{\bsnm{Snijders}, \binits{J.G.}}:
\batitle{Relativistic regular two‐component hamiltonians}.
\bjtitle{The Journal of Chemical Physics}
\bvolume{99}(\bissue{6}),
\bfpage{4597}--\blpage{4610}
(\byear{1993})
\doiurl{10.1063/1.466059}
\end{barticle}
\endbibitem

\bibitem[\protect\citeauthoryear{Tkatchenko and Scheffler}{2009}]{tkatchenko_accurate_2009}
\begin{barticle}
\bauthor{\bsnm{Tkatchenko}, \binits{A.}},
\bauthor{\bsnm{Scheffler}, \binits{M.}}:
\batitle{Accurate {Molecular} {Van} {Der} {Waals} {Interactions} from {Ground}-{State} {Electron} {Density} and {Free}-{Atom} {Reference} {Data}}.
\bjtitle{Physical Review Letters}
\bvolume{102}(\bissue{7}),
\bfpage{073005}
(\byear{2009})
\doiurl{10.1103/PhysRevLett.102.073005}
\end{barticle}
\endbibitem

\bibitem[\protect\citeauthoryear{Nocedal and Wright}{2006}]{nocedal_numerical_2006}
\begin{bbook}
\bauthor{\bsnm{Nocedal}, \binits{J.}},
\bauthor{\bsnm{Wright}, \binits{S.J.}}:
\bbtitle{Numerical {Optimization}}.
\bsertitle{Springer {Series} in {Operations} {Research} and {Financial} {Engineering}}.
\bpublisher{Springer}, \blocation{???}
(\byear{2006}).
\doiurl{10.1007/978-0-387-40065-5} .
\burl{http://link.springer.com/10.1007/978-0-387-40065-5}
\end{bbook}
\endbibitem

\bibitem[\protect\citeauthoryear{Larsen et~al.}{2017}]{ase-paper}
\begin{barticle}
\bauthor{\bsnm{Larsen}, \binits{A.H.}},
\bauthor{\bsnm{Mortensen}, \binits{J.J.}},
\bauthor{\bsnm{Blomqvist}, \binits{J.}},
\bauthor{\bsnm{Castelli}, \binits{I.E.}},
\bauthor{\bsnm{Christensen}, \binits{R.}},
\bauthor{\bsnm{Dułak}, \binits{M.}},
\bauthor{\bsnm{Friis}, \binits{J.}},
\bauthor{\bsnm{Groves}, \binits{M.N.}},
\bauthor{\bsnm{Hammer}, \binits{B.}},
\bauthor{\bsnm{Hargus}, \binits{C.}},
\bauthor{\bsnm{Hermes}, \binits{E.D.}},
\bauthor{\bsnm{Jennings}, \binits{P.C.}},
\bauthor{\bsnm{Jensen}, \binits{P.B.}},
\bauthor{\bsnm{Kermode}, \binits{J.}},
\bauthor{\bsnm{Kitchin}, \binits{J.R.}},
\bauthor{\bsnm{Kolsbjerg}, \binits{E.L.}},
\bauthor{\bsnm{Kubal}, \binits{J.}},
\bauthor{\bsnm{Kaasbjerg}, \binits{K.}},
\bauthor{\bsnm{Lysgaard}, \binits{S.}},
\bauthor{\bsnm{Maronsson}, \binits{J.B.}},
\bauthor{\bsnm{Maxson}, \binits{T.}},
\bauthor{\bsnm{Olsen}, \binits{T.}},
\bauthor{\bsnm{Pastewka}, \binits{L.}},
\bauthor{\bsnm{Peterson}, \binits{A.}},
\bauthor{\bsnm{Rostgaard}, \binits{C.}},
\bauthor{\bsnm{Schiøtz}, \binits{J.}},
\bauthor{\bsnm{Schütt}, \binits{O.}},
\bauthor{\bsnm{Strange}, \binits{M.}},
\bauthor{\bsnm{Thygesen}, \binits{K.S.}},
\bauthor{\bsnm{Vegge}, \binits{T.}},
\bauthor{\bsnm{Vilhelmsen}, \binits{L.}},
\bauthor{\bsnm{Walter}, \binits{M.}},
\bauthor{\bsnm{Zeng}, \binits{Z.}},
\bauthor{\bsnm{Jacobsen}, \binits{K.W.}}:
\batitle{The atomic simulation environment—a python library for working with atoms}.
\bjtitle{Journal of Physics: Condensed Matter}
\bvolume{29}(\bissue{27}),
\bfpage{273002}
(\byear{2017})
\end{barticle}
\endbibitem

\bibitem[\protect\citeauthoryear{Ramakrishnan et~al.}{2014}]{ramakrishnan_quantum_2014}
\begin{barticle}
\bauthor{\bsnm{Ramakrishnan}, \binits{R.}},
\bauthor{\bsnm{Dral}, \binits{P.O.}},
\bauthor{\bsnm{Rupp}, \binits{M.}},
\bauthor{\bsnm{Lilienfeld}, \binits{O.A.}}:
\batitle{Quantum chemistry structures and properties of 134 kilo molecules}.
\bjtitle{Scientific Data}
\bvolume{1}(\bissue{1}),
\bfpage{140022}
(\byear{2014})
\doiurl{10.1038/sdata.2014.22}
\end{barticle}
\endbibitem

\bibitem[\protect\citeauthoryear{Bennett et~al.}{2000}]{bennett2000constrained}
\begin{botherref}
\oauthor{\bsnm{Bennett}, \binits{K.P.}},
\oauthor{\bsnm{Bradley}, \binits{P.S.}},
\oauthor{\bsnm{Demiriz}, \binits{A.}}:
Constrained k-means clustering.
Technical Report MSR-TR-2000-65,
Microsoft Research
(May 2000).
\url{https://www.microsoft.com/en-us/research/publication/constrained-k-means-clustering/}
\end{botherref}
\endbibitem

\bibitem[\protect\citeauthoryear{Berendsen et~al.}{1984}]{berendsen_molecular_1984}
\begin{barticle}
\bauthor{\bsnm{Berendsen}, \binits{H.J.C.}},
\bauthor{\bsnm{Postma}, \binits{J.P.M.}},
\bauthor{\bsnm{Gunsteren}, \binits{W.F.}},
\bauthor{\bsnm{DiNola}, \binits{A.}},
\bauthor{\bsnm{Haak}, \binits{J.R.}}:
\batitle{Molecular dynamics with coupling to an external bath}.
\bjtitle{The Journal of Chemical Physics}
\bvolume{81}(\bissue{8}),
\bfpage{3684}--\blpage{3690}
(\byear{1984})
\doiurl{10.1063/1.448118}
\end{barticle}
\endbibitem

\bibitem[\protect\citeauthoryear{Nosé}{1984}]{nose_unified_1984}
\begin{barticle}
\bauthor{\bsnm{Nosé}, \binits{S.}}:
\batitle{A unified formulation of the constant temperature molecular dynamics methods}.
\bjtitle{The Journal of Chemical Physics}
\bvolume{81}(\bissue{1}),
\bfpage{511}--\blpage{519}
(\byear{1984})
\doiurl{10.1063/1.447334}
\end{barticle}
\endbibitem

\bibitem[\protect\citeauthoryear{Hoover}{1985}]{hoover_canonical_1985}
\begin{barticle}
\bauthor{\bsnm{Hoover}, \binits{W.G.}}:
\batitle{Canonical dynamics: {Equilibrium} phase-space distributions}.
\bjtitle{Physical Review A}
\bvolume{31}(\bissue{3}),
\bfpage{1695}--\blpage{1697}
(\byear{1985})
\doiurl{10.1103/PhysRevA.31.1695}
\end{barticle}
\endbibitem

\bibitem[\protect\citeauthoryear{Ceriotti et~al.}{2009}]{ceriotti_langevin_2009}
\begin{barticle}
\bauthor{\bsnm{Ceriotti}, \binits{M.}},
\bauthor{\bsnm{Bussi}, \binits{G.}},
\bauthor{\bsnm{Parrinello}, \binits{M.}}:
\batitle{Langevin {Equation} with {Colored} {Noise} for {Constant}-{Temperature} {Molecular} {Dynamics} {Simulations}}.
\bjtitle{Physical Review Letters}
\bvolume{102}(\bissue{2}),
\bfpage{020601}
(\byear{2009})
\doiurl{10.1103/PhysRevLett.102.020601}
\end{barticle}
\endbibitem

\bibitem[\protect\citeauthoryear{Wiener}{1930}]{wiener_generalized_1930}
\begin{barticle}
\bauthor{\bsnm{Wiener}, \binits{N.}}:
\batitle{Generalized harmonic analysis}.
\bjtitle{Acta Mathematica}
\bvolume{55}(\bissue{none}),
\bfpage{117}--\blpage{258}
(\byear{1930})
\doiurl{10.1007/BF02546511}
\end{barticle}
\endbibitem

\bibitem[\protect\citeauthoryear{Blackman and Tukey}{1958}]{blackman_measurement_1958}
\begin{barticle}
\bauthor{\bsnm{Blackman}, \binits{R.B.}},
\bauthor{\bsnm{Tukey}, \binits{J.W.}}:
\batitle{The measurement of power spectra from the point of view of communications engineering — {Part} {I}}.
\bjtitle{The Bell System Technical Journal}
\bvolume{37}(\bissue{1}),
\bfpage{185}--\blpage{282}
(\byear{1958})
\doiurl{10.1002/j.1538-7305.1958.tb03874.x}
\end{barticle}
\endbibitem

\bibitem[\protect\citeauthoryear{Schütt et~al.}{2019}]{schutt_schnetpack_2019}
\begin{barticle}
\bauthor{\bsnm{Schütt}, \binits{K.T.}},
\bauthor{\bsnm{Kessel}, \binits{P.}},
\bauthor{\bsnm{Gastegger}, \binits{M.}},
\bauthor{\bsnm{Nicoli}, \binits{K.A.}},
\bauthor{\bsnm{Tkatchenko}, \binits{A.}},
\bauthor{\bsnm{Müller}, \binits{K.-R.}}:
\batitle{{SchNetPack}: {A} {Deep} {Learning} {Toolbox} {For} {Atomistic} {Systems}}.
\bjtitle{Journal of Chemical Theory and Computation}
\bvolume{15}(\bissue{1}),
\bfpage{448}--\blpage{455}
(\byear{2019})
\doiurl{10.1021/acs.jctc.8b00908}
\end{barticle}
\endbibitem

\bibitem[\protect\citeauthoryear{Pedregosa et~al.}{2011}]{scikit-learn}
\begin{barticle}
\bauthor{\bsnm{Pedregosa}, \binits{F.}},
\bauthor{\bsnm{Varoquaux}, \binits{G.}},
\bauthor{\bsnm{Gramfort}, \binits{A.}},
\bauthor{\bsnm{Michel}, \binits{V.}},
\bauthor{\bsnm{Thirion}, \binits{B.}},
\bauthor{\bsnm{Grisel}, \binits{O.}},
\bauthor{\bsnm{Blondel}, \binits{M.}},
\bauthor{\bsnm{Prettenhofer}, \binits{P.}},
\bauthor{\bsnm{Weiss}, \binits{R.}},
\bauthor{\bsnm{Dubourg}, \binits{V.}},
\bauthor{\bsnm{Vanderplas}, \binits{J.}},
\bauthor{\bsnm{Passos}, \binits{A.}},
\bauthor{\bsnm{Cournapeau}, \binits{D.}},
\bauthor{\bsnm{Brucher}, \binits{M.}},
\bauthor{\bsnm{Perrot}, \binits{M.}},
\bauthor{\bsnm{Duchesnay}, \binits{E.}}:
\batitle{Scikit-learn: Machine learning in {P}ython}.
\bjtitle{Journal of Machine Learning Research}
\bvolume{12},
\bfpage{2825}--\blpage{2830}
(\byear{2011})
\end{barticle}
\endbibitem

\bibitem[\protect\citeauthoryear{Henschel et~al.}{2020}]{henschel_theoretical_2020}
\begin{barticle}
\bauthor{\bsnm{Henschel}, \binits{H.}},
\bauthor{\bsnm{Andersson}, \binits{A.T.}},
\bauthor{\bsnm{Jespers}, \binits{W.}},
\bauthor{\bsnm{Mehdi~Ghahremanpour}, \binits{M.}},
\bauthor{\bsnm{Spoel}, \binits{D.}}:
\batitle{Theoretical {Infrared} {Spectra}: {Quantitative} {Similarity} {Measures} and {Force} {Fields}}.
\bjtitle{Journal of Chemical Theory and Computation}
\bvolume{16}(\bissue{5}),
\bfpage{3307}--\blpage{3315}
(\byear{2020})
\doiurl{10.1021/acs.jctc.0c00126}
\end{barticle}
\endbibitem

\bibitem[\protect\citeauthoryear{Pracht et~al.}{2020}]{pracht_comprehensive_2020}
\begin{barticle}
\bauthor{\bsnm{Pracht}, \binits{P.}},
\bauthor{\bsnm{Grant}, \binits{D.F.}},
\bauthor{\bsnm{Grimme}, \binits{S.}}:
\batitle{Comprehensive {Assessment} of {GFN} {Tight}-{Binding} and {Composite} {Density} {Functional} {Theory} {Methods} for {Calculating} {Gas}-{Phase} {Infrared} {Spectra}}.
\bjtitle{Journal of Chemical Theory and Computation}
\bvolume{16}(\bissue{11}),
\bfpage{7044}--\blpage{7060}
(\byear{2020})
\doiurl{10.1021/acs.jctc.0c00877}
\end{barticle}
\endbibitem

\bibitem[\protect\citeauthoryear{Rubner et~al.}{2000}]{rubner_earth_2000}
\begin{barticle}
\bauthor{\bsnm{Rubner}, \binits{Y.}},
\bauthor{\bsnm{Tomasi}, \binits{C.}},
\bauthor{\bsnm{Guibas}, \binits{L.J.}}:
\batitle{The {Earth} {Mover}'s {Distance} as a {Metric} for {Image} {Retrieval}}.
\bjtitle{International Journal of Computer Vision}
\bvolume{40}(\bissue{2}),
\bfpage{99}--\blpage{121}
(\byear{2000})
\doiurl{10.1023/A:1026543900054}
\end{barticle}
\endbibitem

\bibitem[\protect\citeauthoryear{Himanen et~al.}{2019}]{Himanen/Geurts/Foster/Rinke:2019}
\begin{barticle}
\bauthor{\bsnm{Himanen}, \binits{L.}},
\bauthor{\bsnm{Geurts}, \binits{A.}},
\bauthor{\bsnm{Foster}, \binits{A.S.}},
\bauthor{\bsnm{Rinke}, \binits{P.}}:
\batitle{Data-driven materials science: Status, challenges, and perspectives}.
\bjtitle{Adv. Sci.}
\bvolume{6}(\bissue{21}),
\bfpage{1900808}
(\byear{2019})
\end{barticle}
\endbibitem

\end{thebibliography}

\end{document}